\documentclass{ametsoc}

\journal{jpo}

\bibpunct{(}{)}{;}{a}{}{,}

\title{Near-inertial wave critical layers over sloping bathymetry}

\authors{Lixin Qu\correspondingauthor{Lixin Qu, Department of Earth System Science, Stanford University, 473 Via Ortega, Stanford, CA. USA} and Leif N. Thomas }

\affiliation{Department of Earth System Science, Stanford University}

\email{lixinqu@stanford.edu}

\extraauthor{Robert D. Hetland}
\extraaffil{Department of Oceanography, Texas A\&M University}

\abstract{This study describes a specific type of critical layer for near-inertial waves (NIWs) that forms when isopycnals run parallel to sloping bathymetry. Upon entering this slantwise critical layer, the group velocity of the waves decreases to zero and the NIWs become trapped and amplified, which can enhance mixing. A realistic simulation of anticyclonic eddies on the Texas-Louisiana shelf reveals that such critical layers can form where the eddies impinge onto the sloping bottom. Velocity shear bands in the simulation indicate that wind-forced NIWs are radiated downward from the surface in the eddies, bend upward near the bottom, and enter critical layers over the continental shelf, resulting in inertially-modulated enhanced mixing. Idealized simulations designed to capture this flow reproduce the wave propagation and enhanced mixing. The link between the enhanced mixing and  wave trapping in the slantwise critical layer is made using ray-tracing and an analysis of the waves' energetics in the idealized simulations. An ensemble of simulations is performed spanning the relevant parameter space that demonstrates that the strength of the mixing is correlated with the degree to which NIWs are trapped in the critical layers. While the application here is for a shallow coastal setting, the mechanisms could be active in the open ocean as well where isopycnals align with bathymetry.}

\begin{document}

\maketitle

\section{Introduction}
Processes that drive enhanced mixing near the sloping seafloor have received increased attention in recent years due to their potential role in shaping water mass transformation and diapycnal upwelling \citep{ferrari16, mcdougall17, callies18}.  One such process is critical reflection of inertia-gravity waves (IGWs) which occurs when wave rays align with bathymetry such that upon reflection, wave energy is focused near the bottom, leading to bores, boluses, vortices, turbulence, and mixing \citep{cacchione74, kunze04, chalamalla13}. The phenomenon has almost exclusively been studied with the internal tides in mind since they carry a significant fraction of the energy in the oceanic internal wave field and because many continental slopes are near-critical for tidal frequencies \citep{cacchione02}. Near-inertial waves (NIWs) carry a comparable amount of energy and have a power input into them that is similar to the internal tides \citep{alford03, ferrari_wunsch09}, but they have not been considered as key players in driving mixing via critical reflection on sloping topography. It seems reasonable to neglect NIWs in this regard, because according to classical internal wave theory, NIWs propagate at very shallow angles and therefore would only experience critical reflection off nearly-flat bathymetry, which would not result in much wave amplification. However, classical internal wave theory does not account for the modification of wave propagation by background flows. 

In particular, baroclinic, geostrophically-balanced flows can greatly alter the propagation pathways of NIWs, resulting in rays with slopes
\begin{equation}
s_{ray} = s_{\rho} \pm \sqrt{\frac{\omega^2-\omega_{min}^2}{N^2}}
\label{ray_slope}
\end{equation}
(where $\omega$ is the frequency of the wave, $\omega_{min}$ is the minimum frequency allowable for IGWs, and $N^2$ is the square of the buoyancy frequency) that are symmetric about isopycnals of slope $s_{\rho}$, which are tilted in baroclinic flows \citep{mooers1975several, whitt2013near}. The minimum frequency of IGWs tends to be close to the inertial frequency $f$, thus NIWs with $\omega \approx f$ propagate along rays that run nearly parallel to isopycnals, i.e. $s_{ray} \approx s_{\rho}$. This opens the possibility that NIWs can experience critical reflection off sloping bathymetry when isopycnals are aligned with the bottom slope, $\alpha$. However, when NIWs approach a region where their frequency is equal to $\omega_{min}$, their group velocity goes to zero, and rather than reflecting, the waves can be trapped and amplified in critical layers \citep{kunze1985near} which are slantwise if $s_{\rho} \ne 0$ \citep{ whitt2013near}. In this article we will demonstrate that the scenario with $\alpha=s_{\rho}$, can give rise to enhanced near-bottom mixing associated with NIWs entering such slantwise critical layers.  

It is not unusual to find flows in the ocean with isopycnals that follow bathymetry. Dense overflows, such as those found on the western Weddell Sea margin, in the Denmark Strait, or over the Iceland-Faroe Ridge, for example, naturally generate bottom-intensified along-isobath currents where the isopycnals that encapsulate their dense waters blanket topographic features \citep{muench95, girton01, beaird13}. Isopycnals can also be aligned with bathymetry by upslope Ekman flows associated with the Ekman arrest of currents flowing opposite to the direction of Kelvin wave propagation \citep{garrett93}. The Florida Current is an example of such a flow and indeed has isopycnals that tend to parallel the continental slope off of Florida \citep{winkel02}. Wind-forced coastal upwelling can also result in isopycnals paralleling the bottom, and there is evidence that NIWs are amplified in critical layers during periods of upwelling but not during downwelling \citep{federiuk1996model}.

\begin{figure}[t]
\centerline{\includegraphics[width=0.8\textwidth]{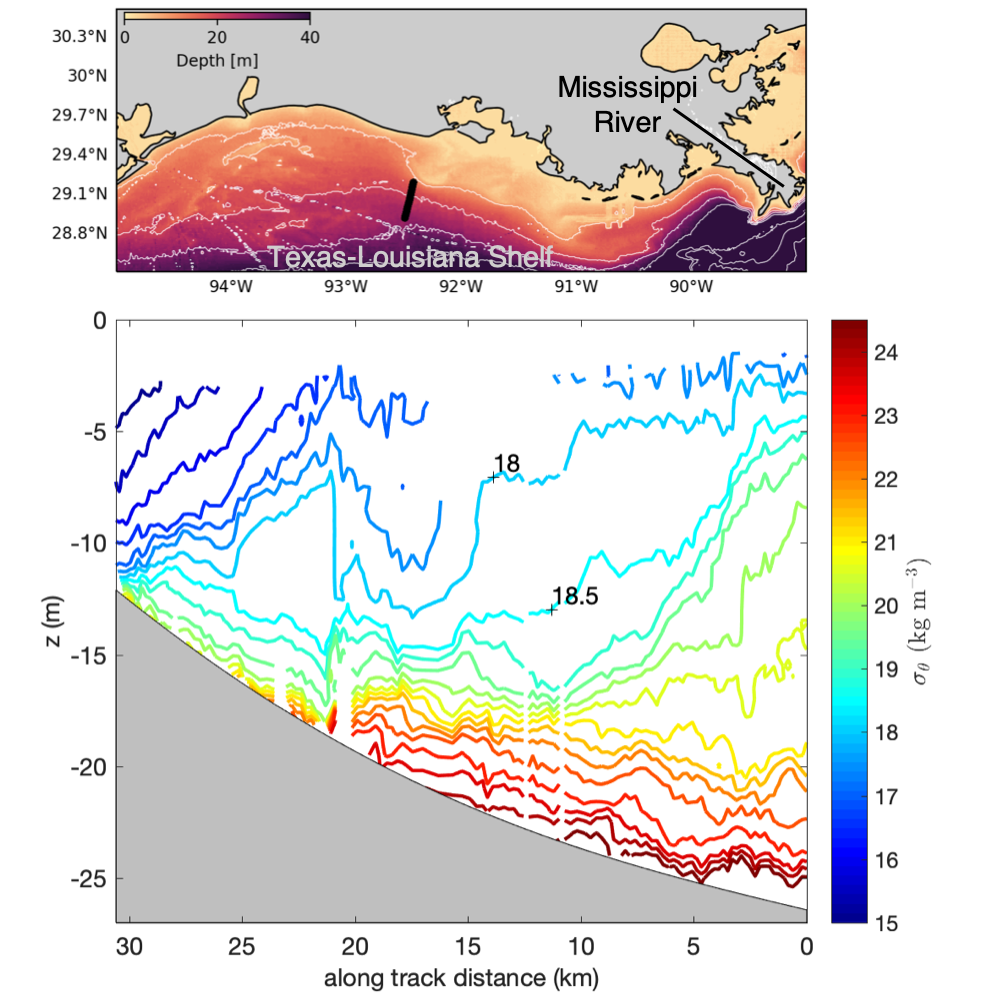}}
\caption{Section of potential density (contoured every 0.5 kg m$^{-3}$) across the Texas-Louisiana shelf (the location of which is indicated by the black line in the upper panel) from observations made on June 14, 2010 as part of the Mechanisms Controlling Hypoxia study.}
\label{MCH_section}
\end{figure}

Another example of flow that can meet the $s_{\rho}=\alpha$ criterion are the currents on the inshore side of the anticyclonic eddies that form in the Mississippi/Atchafalaya river plume on the Texas-Louisiana shelf. High-resolution hydrographic sections made on this shelf as part of the Mechanisms Controlling Hypoxia study \citep[e.g.][]{zhang15} illustrate the structure of the density field associated with these eddies (Fig. \ref{MCH_section}). Density surfaces form a bowl-like structure within the anticyclones while near the bottom isopycnals create a stratified layer that shoals towards the shore with $s_{\rho} \approx \alpha$. 

During summer, the anticyclones on the Texas-Louisiana shelf coincide with strong near-inertial currents driven by the diurnal land-sea breeze which is near-resonant since the diurnal frequency is close to $f$ \citep{zhang2009near}. Therefore if these near-inertial currents create downward propagating waves, then the anticyclones would provide the ideal conditions for critical reflection of NIWs over sloping topography. Realistic simulations of the circulation and wave field on the Texas-Louisiana shelf suggest that these conditions are indeed met. We will describe these simulation in section \ref{TXLA_section} and use them to motivate theoretical analyses (section \ref{theory_section}) and idealized simulations (section \ref{idealized_sims_section}) aimed at understanding the underlying physics behind the phenomenon. With this coastal scenario as an example, the ultimate goal of this study is to build the link between wave trapping within a slantwise critical layer and the enhanced bottom mixing that it can induce.  We will end the article with discussions of the enhanced diapycnal transport in bottom critical layers and the mixing enhancement (section \ref{discussion_section}), which will be followed by a summary of our conclusions (section \ref{conclusions_section}).

\section{Realistic simulations of NIW-eddy interactions on the Texas-Louisiana shelf}\label{TXLA_section}

\begin{figure}[t]
\centerline{\includegraphics[width=\textwidth]{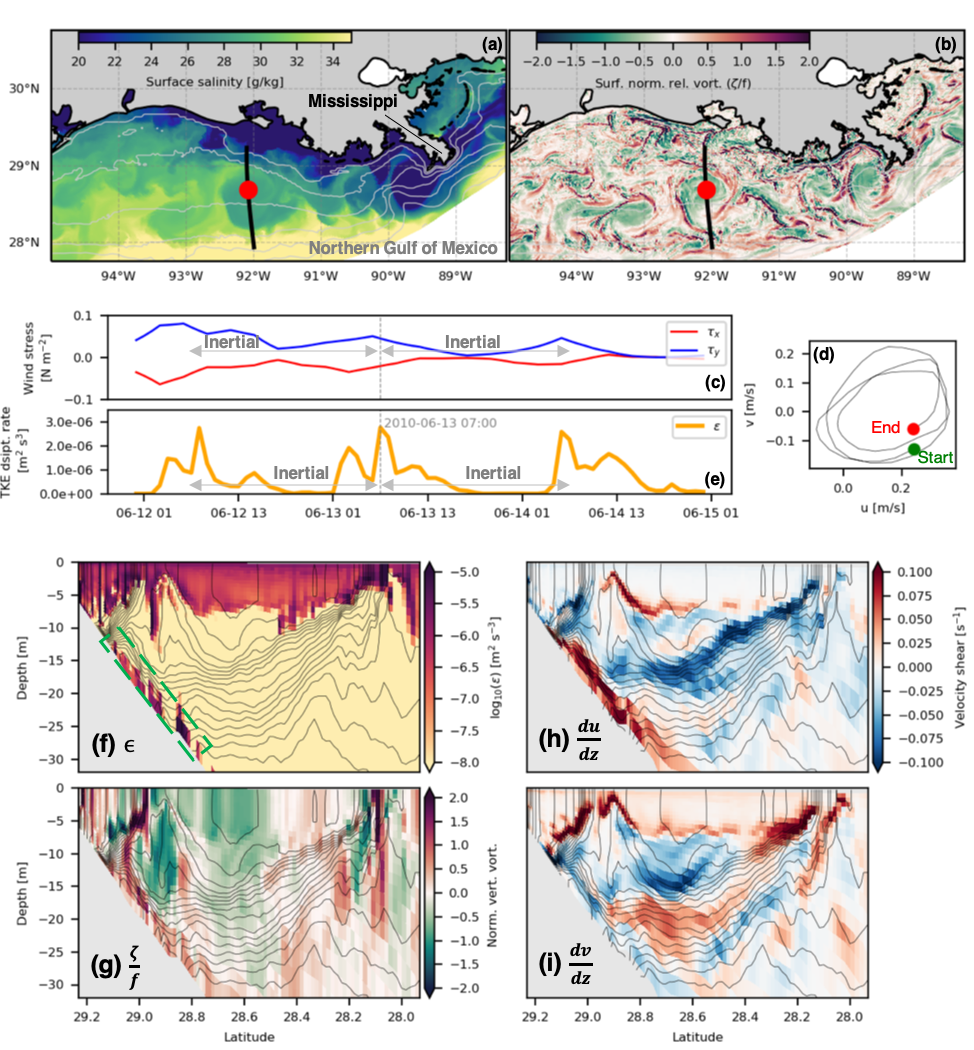}}
\caption{(a and b) Snapshots of surface salinity and normalized relative vorticity $\frac{\zeta}{f}$ from the TXLA simulation. (c) Time series of the zonal component (red) and meridional (blue) component of the wind stress at the red dot marked in (a) and (b). (d) Surface velocity hodograph at the red dot from Jun 12 to 14. (e) Time series of the volume-averaged (in the green dashed box in panel f) TKE dissipation rate $\epsilon$. (f, g, h, and i) Sections of $\epsilon$, $\frac{\zeta}{f}$, $\frac{du}{dz}$, and $\frac{dv}{dz}$ along the black line marked in (a) and (b). The time of the snapshots is 7:00, Jun 13, 2010, indicated by the dashed grey line (c) and (e).}
\label{fig1}
\end{figure}

Here, we present results from the TXLA model, a realistic simulation on the Texas-Louisiana shelf in the northern Gulf of Mexico, that highlights the interaction of NIWs with anticyclones in a coastal region with sloping bathymetry \citep{zhang2012high}. In the northern Gulf, the Mississippi and Atchafalaya rivers create a large region of buoyant, relatively fresh water over the Texas-Louisiana shelf. The river plume front is unstable to baroclinic inabilities during summertime, due to a pooling of fresh water over the Louisiana shelf by weak upwelling winds and a lack of storm fronts, which generates a rich field of eddies \citep{hetland2017suppression,qu2020non}. As illustrated in the TXLA model output, the eddies are characteristically fresh (buoyant) anti-cyclones, surrounded by strong cyclonic filaments at their edges (Fig.~\ref{fig1}a and \ref{fig1}b). In addition, since storms are infrequent in the summertime and winds are generally mild, the diurnal land-sea breeze becomes an important forcing mechanism (Fig.~\ref{fig1}c). Noting that this region is near the critical latitude, $29^{\circ}$ N, the diurnal land-sea breeze is nearly resonant with the local inertial frequency, such that the land-sea breeze drives significant near-inertial oscillations, with peak clockwise rotating velocities of around 0.5 m~s$^{-1}$ (Fig.~\ref{fig1}d). There are indications that these oscillations at the surface become downward propagating NIWs that radiate away from the offshore edge of anticyclones towards the shoaling bathymetry. Namely, bands of vertical shear in the zonal and meridional velocities descend from the offshore edge of the eddy and bend upwards with isopycnals near the bottom on the inshore side of the eddy (Fig.~\ref{fig1}h and \ref{fig1}i). The shear bands propagate upward (not shown) indicating upward phase propagation and hence suggesting downward energy propagation. 

Interestingly, the turbulent kinetic energy (TKE) dissipation is enhanced near the bottom where the waves are approaching (marked by the green box in Fig.~\ref{fig1}f) with values that are comparable to the dissipation near the surface. The TKE dissipation rate $\epsilon$ is diagnosed via the $k-\epsilon$ turbulence closure scheme. The bottom dissipation pulses over an inertial period (Fig.~\ref{fig1}e) indicating a relationship between the enhanced dissipation and the resonantly forced near-inertial motions. We explore the underlying physics behind this relationship using theory and idealized simulations in the next two sections.

\section{Theory}\label{theory_section}
\begin{figure}[t]
\centerline{\includegraphics[width=0.9\textwidth]{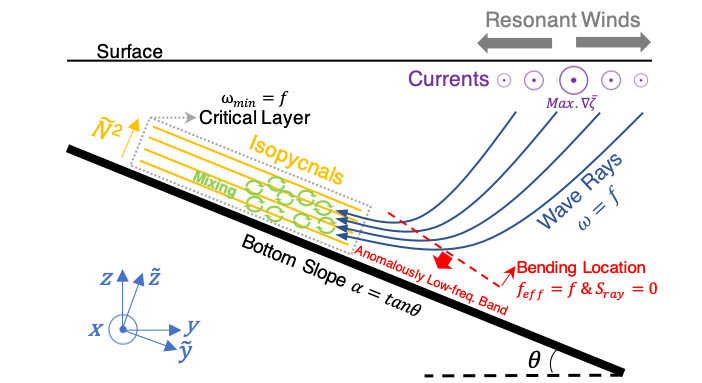}}
\caption{Schematic of the theoretical model. Trapping and amplification of inertial waves within the slantwise critical layer formed when isopycnals run parallel to the bottom slope results in enhanced mixing.}
\label{fig2}
\end{figure}

In this section, we develop a simple theoretical model to interpret and link the three key features revealed by the TXLA simulation: 1) downward propagation of near-inertial energy from the surface; 2) upward bending of shear bands near the bottom; 3) enhanced dissipation in the stratified layer over the bottom. The theoretical model that integrates these key elements is schematized in Fig.~\ref{fig2} and is elaborated on below. 

\subsection{Downward propagation of near-inertial wave from the surface}
In our theoretical model, the wind oscillates at the local inertial frequency (such as the diurnal land-sea breeze at the latitude of 29$^\circ$), so inertial waves with $\omega=f$ are resonantly forced. In the absence of a background flow, the minimum frequency of IGWs is $f$, therefore the slope of rays, (\ref{ray_slope}), for these inertial waves is zero (since isopycnals are flat when there are no currents), and wave energy cannot propagate vertically.  In the presence of a background flow, $\overline{u}$, that follows the thermal wind balance: $M^2=f \partial \overline{u}/\partial z=-\partial \overline{b}/\partial y$, and that has a vertical vorticity $\overline{\zeta}=-\partial \overline{u}/\partial y$, the minimum frequency of IGWs is
\begin{equation}
    \omega_{min} = \sqrt{f^2_{eff}-M^4/N^2}.
 \label{eq:min_omega}
\end{equation}
where $f_{eff}=\sqrt{f(f+\overline{\zeta})}$ is the effective inertial frequency \citep{mooers1975several,whitt2013near}. Consequently, in regions of anticyclonic vorticity $\omega_{min}<f$ and therefore inertial waves have rays with non-zero slopes, allowing for vertical wave propagation. This results in enhanced downward energy propagation of NIWs in anticyclones, a phenomenon that is known as the "inertial chimney" effect   \citep{lee1998inertial}.

However, to vertically propagate, NIWs need to acquire a finite horizontal wavelength and a non-zero horizontal wavenumber. The horizontal wavelength of NIWs can be reduced due to the presence of vorticity gradients, via the process of refraction \citep{young1997propagation, asselin2020penetration}.  Gradients in $\overline{\zeta}$ set up lateral differences in wave phase since near-inertial oscillations separated a short distance from one another oscillate at slightly different frequencies. As a result, the near-inertial motions develop a horizontal wavenumber whose magnitude increases linearly with time at a rate that is proportional to the gradient in $f_{eff}$ \citep{van_meurs98}. As illustrated in Fig.~\ref{fig2}, we envision that such refraction is active at the offshore edge of the anticyclone near the maximum in velocity where the vorticity gradient is maximum and $f_{eff}=f$. Therefore it is here where resonantly-forced inertial waves will develop a horizontal wavenumber and radiate down into the anticyclone.

\subsection{Reversal of vertical energy propagation in the anomalously low-frequency regime}

The upward bending of the shear bands at depth on the inshore side of the anticyclone seen in the TXLA simulation (e.g. Fig.~\ref{fig1}h and \ref{fig1}i) suggests that the vertical propagation of the surface-generated NIWs changes sign at depths well above the bottom. Such a reversal of vertical energy propagation not due to bottom reflections is possible in background flows with baroclinicity. This follows from the expression for the slope of wave rays (\ref{ray_slope}) which can change sign without switching characteristics (that is, without switching roots in (\ref{ray_slope}), which occurs at reflections). The vertical direction of energy propagation reverses sign where $s_{ray}=0$, which occurs where the wave's frequency is equal to the local effective inertial frequency $\omega=f_{eff}$. In a flow that is baroclinic, since $\omega_{min}<f_{eff}$, the waves can propagate past this location and when they do so, $s_{ray}$ and the vertical component of their group velocity changes sign. In this region, the wave's frequency is less than $f_{eff}$ but greater than $\omega_{min}$, i.e. $\omega_{min}<\omega<f_{eff}$. This is the so-called anomalously low-frequency regime defined by \cite{mooers1975several}, where NIWs are characterized by unusual behavior. In particular, the vertical components of the group and phase velocities can be in the same direction in the anomalously low-frequency regime \citep{whitt2013near}. This is observed in the TXLA simulation, since the shear bands that bend upwards near the bottom on the inshore side of the eddy (thus fluxing energy to the shallows) also propagate upwards in time, indicating a positive phase velocity. 

This process is schematized in Fig.~\ref{fig2} for inertial waves. The location where the wave rays start to bend is where $s_{ray}=0$ and $f_{eff}=f$. After passing the bending location, $s_{ray}$ increases from zero so that the wave rays bend upwards. At the same time, the waves enter the anomalously low-frequency regime (where $f_{eff}>f$), and according to the theory, the phase velocity should have the same sign as the group velocity such that the phase also propagate upwards.

\subsection{Trapping in a slantwise critical layer}
As NIWs enter the anomalously-low frequency regime, their frequency approaches $\omega_{min}$. At the location or locations where $\omega=\omega_{min}$, the magnitude of the group velocity
\begin{equation}
|\mathbf{c}_g|=\frac{N^2}{\omega |m|} \sqrt{\frac{(\omega^2-\omega_{min}^2)(1+s_{ray}^2)}{N^2}} 
\label{c_g}
\end{equation}
($m$ is the wave's vertical wavenumber) goes to zero \citep{whitt2013near}. These locations can be either turning points or critical layers depending on the geometry of the contour where $\omega_{min}=\omega$. This contour is known as the separatrix and if it is aligned with wave rays, waves cannot radiate away from this boundary and are trapped, thus forming a critical layer. From (\ref{ray_slope}), wave rays run parallel to isopycnals at the separatix since $\omega=\omega_{min}$ there, hence an alignment of the separatrix with isopycnals marks the locations of critical layers. In weakly baroclinic anticyclones for example, critical layers are nearly flat and form at the base of the vortices where the vorticity increases with depth  \citep{kunze1985near}. In strongly baroclinic currents in contrast, NIW critical layers tilt with isopycnals and tend to be found in regions of cyclonic vorticity \citep{whitt2013near}. Such slantwise critical layers can form in stratified layers over sloping bathymetry, as we demonstrate below.

Here, we introduce a specific type of slantwise critical layer for inertial waves with $\omega=f$ over sloping bathymetry. This critical layer forms in a stratified layer with isopycnals that run parallel to bathymetry, as illustrated in Fig.~\ref{fig2}, mimicking the layers that have been seen in the observations and simulations of the flows on the Texas-Louisiana shelf (Fig.~\ref{MCH_section} and \ref{fig1}). The tilted isopycnals induce a horizontal buoyancy gradient and, assuming that the flow is in geostrophic balance, create a thermal wind shear  
\begin{equation}
 \begin{aligned}
    \frac{\partial  \overline{u}}{\partial z} = - \frac{1}{f} \frac{\partial  \overline{b}}{\partial y} =- \frac{\Tilde{N}^2}{f} sin \theta \ ,\\
 \end{aligned}
 \label{eq:TW}
\end{equation}
where $\Tilde{N}^2 (\Tilde{z}) \equiv \frac{\partial  \overline{b}}{\partial \Tilde{z}} (\Tilde{z})$ is the "stratification" in the rotated coordinates ($\Tilde{z}$ denotes the slope-normal direction) and $\theta$ is the bottom slope angle (see Appendix A for the derivation). Note that $\Tilde{N}^2 (\Tilde{z})$ only depends on the slope-normal distance $\Tilde{z}$ since the isopycnals are parallel with the sloping bathymetry. The formation of this critical layer is based on the assumption that the bottom boundary layer is an "arrested" Ekman layer, where the along-shore flow reaches a thermal wind balance and that satisfies the no-slip bottom boundary condition causing the across-shore flow to vanish \citep{maccready1993slippery,wenegrat2018submesoscale}. This thermal wind shear induces a finite Richardson number,
\begin{equation}
    Ri_g = \frac{\partial  \overline{b}}{\partial z} \left(\frac{\partial  \overline{u}}{\partial z}\right)^{-2} = \frac{f^2}{\Tilde{N}^2 sin \theta\ tan \theta}   \ ,\\
 \label{eq:Ri}
\end{equation}
where the vertical buoyancy gradient $\frac{\partial  \overline{b}}{\partial z}$ is related to $\Tilde{N}^2$ by $\frac{\partial  \overline{b}}{\partial z} = \Tilde{N}^2 cos \theta$ (see Appendix A). Furthermore, the geostrophic flow is also horizontally sheared because of the sloping bathymetry. The horizontal shear in $\overline{u}$ can be written as 
\begin{equation}
 \begin{aligned}
    \frac{\partial  \overline{u}}{\partial y} =- \frac{\Tilde{N}^2}{f} sin \theta\ tan \theta \ .\\
 \end{aligned}
 \label{eq:uhshear}
\end{equation}
(see Appendix A), yielding a cyclonic vorticity and hence a positive Rossby number
\begin{equation}
    Ro_g = \overline{\zeta}/f = -\frac{\partial  \overline{u}}{\partial y} /f = \frac{\Tilde{N}^2}{f^2} sin \theta\ tan \theta  \ .\\
 \label{eq:Ro}
\end{equation}
In this stratified layer the minimum frequency of IGWs is exactly inertial:
\begin{equation}
    \omega_{min} = \sqrt{f_{eff}^2 - M^4/N^2} = f \sqrt{1+Ro_g-Ri_g^{-1}} =f,
 \label{eq:omin}
\end{equation}
since the contributions from vorticity (\ref{eq:Ro}) and baroclinicity (\ref{eq:Ri}) cancel. The minimum frequency can also be expressed in terms of the Ertel potential vorticity (PV) of the geostrophic flow, $\overline{q}=(f+\overline{\zeta})N^2-M^4/f$, i.e. $\omega_{min}=\sqrt{f\overline{q}/N^2}$ \citep{whitt2013near}, indicating that when $\omega_{min}=f$, the PV is equal to the planetary PV, $fN^2$. This provides some insights for how a geostrophic flow with $Ro_g=Ri_g^{-1}$ might form. Namely, it could be generated through advective processes that conserve PV, and be initiated from a flow with low Rossby and inverse Richardson numbers where the PV is approximately equal to the planetary PV. Frontogenesis is one process that is capable of generating such a flow \citep[e.g.][]{pedlosky87}, and upwelling dense water up a slope while maintaining geostrophy could be another possibility.   

Under these conditions, while the minimum frequency is inertial, $\omega_{min}=f$, the effective inertial frequency $f_{eff}=f\sqrt{1+Ro_g}$ is superinertial. Hence inertial waves in the stratified layer enter the anomalously low-frequency regime as $\omega_{min}=f<f_{eff}$. Moreover, since $\omega_{min}$ is uniform and equal to $f$ in this layer, the separatrix runs parallel to isopycnals and the criterion for a slantwise critical layer is met.  Finally, from (\ref{c_g}) it is clear that the group velocity is equal to zero in the layer and should cause inertial waves to be trapped and amplified there, which could drive enhanced mixing. The enhanced bottom dissipation in the slantwise stratified layer exhibited in the TXLA simulation (Fig.~\ref{fig1}f) suggests that this mechanism is active there. We test these theoretical ideas in a more controlled environment than the TXLA model using idealized simulations, as described in the next section.

\section{Idealized simulations}\label{idealized_sims_section}
\subsection{Base run}
The Regional Ocean Modeling System (ROMS) is employed in this study, which is a free-surface, hydrostatic, primitive equation ocean model that uses an S-coordinate in the vertical direction \citep{shchepetkin2005regional}. ROMS is configured to conduct idealized simulations. The model domain represents an idealized coastal region over a continental shelf with a constant slope $\alpha=5\times 10^{-4}$, and with the depths ranging from 5 m to 118 m. The domain has an across-shore span of 226 km and an along-shore width of 4 km. The domain is set to be extremely narrow in the along-shore direction with few grid points so that the variation in the along-shore direction can be assumed to be negligibly small. The horizontal resolution is $220\ m \times 220\ m$. There are 64 layers in the vertical direction with the stretching parameters of $\theta_S=3.0$ and $\theta_B=0.4$. The along-shore boundary conditions are set to be periodic, and the offshore open boundary has a sponge layer that damps the waves propagating towards the open boundary. The Coriolis parameter is equal to the diurnal frequency, i.e., $f=\frac{2\pi}{86400}\ s^{-1} \approx 7.27 \times 10 ^{-5}\ s^{-1}$. The wind forcing is set to mimic a diurnal land-sea breeze - a rectilinear oscillating wind oriented in the across-shore direction with an amplitude of $4\times 10^{-2} N m^{-2}$. The simulation is run for 10 days.

\begin{figure}[t]
\centerline{\includegraphics[width=0.75\textwidth]{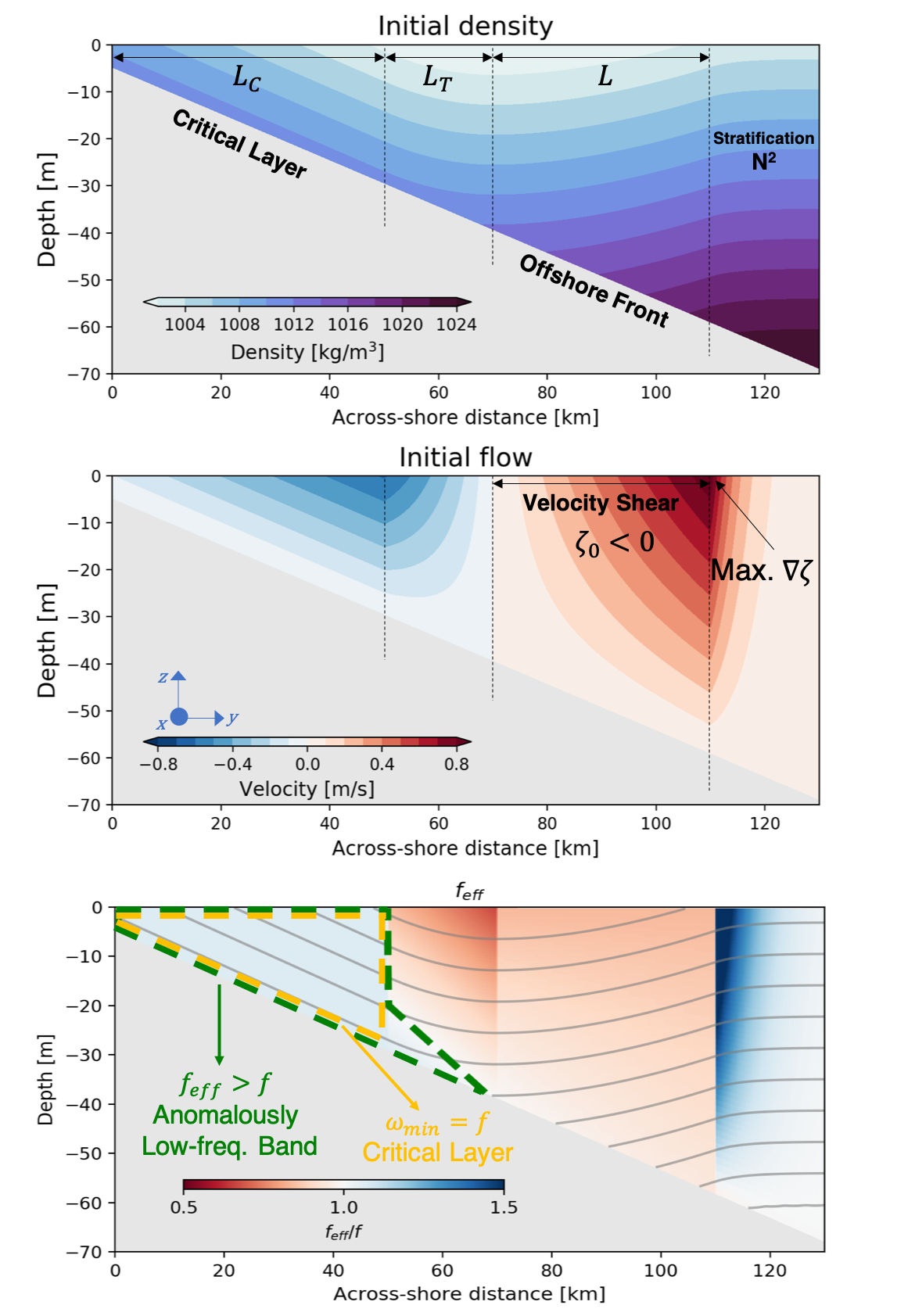}}
\caption{Initial conditions of the density (upper panel) and the along-shore velocity (middle panel) in the base run. The parameters of the base run are listed in Tab.~\ref{tab1}. Initial distribution of $f_{eff}$ (lower panel). The boundaries of the anomalously low-frequency regime and the critical layer are marked by the green dashed lines and the orange lines, respectively.}
\label{fig3}
\end{figure}

The initial conditions correspond to an anticyclonic baroclinic flow with a slantwise critical layer onshore and a buoyant front offshore (Fig.~\ref{fig3}), with parameters that are based on the realistic simulation. The critical layer has an across-shore width of $L_C=50$ km, and the offshore front has a width of $L=40$ km. Note that there is a transition zone with a width of $L_T=20$ km in the middle where the horizontal buoyancy gradient linearly decreases to zero with increasing across-shore distance. The flow and density fields have no variations in the along-shore direction. The stratification is set to $N^2 = 3 \times 10^{-3} s^{-2}$, a value based on the realistic simulation, and is constant across the domain. The density structure of the critical layer is determined by $N^2$ and $\alpha$, and the flow is initially in a thermal wind balance with the density field. The density structure of the offshore buoyant front is determined by the velocity structure due to the constraint of the thermal wind balance.  In the horizontal direction, moving in the across-shore direction, the surface velocity at the offshore front increases from zero with a vorticity of $ \zeta_0 = -0.3f$ between $L_C+L_T \le y \le L_C+L_T+L$ and decays exponentially to zero offshore and outside of this region. In the vertical direction, the velocity decays linearly to zero towards the bottom. There is no initial across-shore flow in the domain. The MPDATA scheme is used for the tracer advection \citep{smolarkiewicz1998mpdata}. The $k-\epsilon$ turbulence closure scheme is used to parameterize vertical mixing, and the Canuto A stability function formulation is applied \citep{umlauf2003generic,canuto2001ocean}. No explicit lateral diffusivity is used in the simulation. The parameters used to configure this simulation are listed in the first row of Tab.~\ref{tab1}.

\begin{figure}[t]
\centerline{\includegraphics[width=\textwidth]{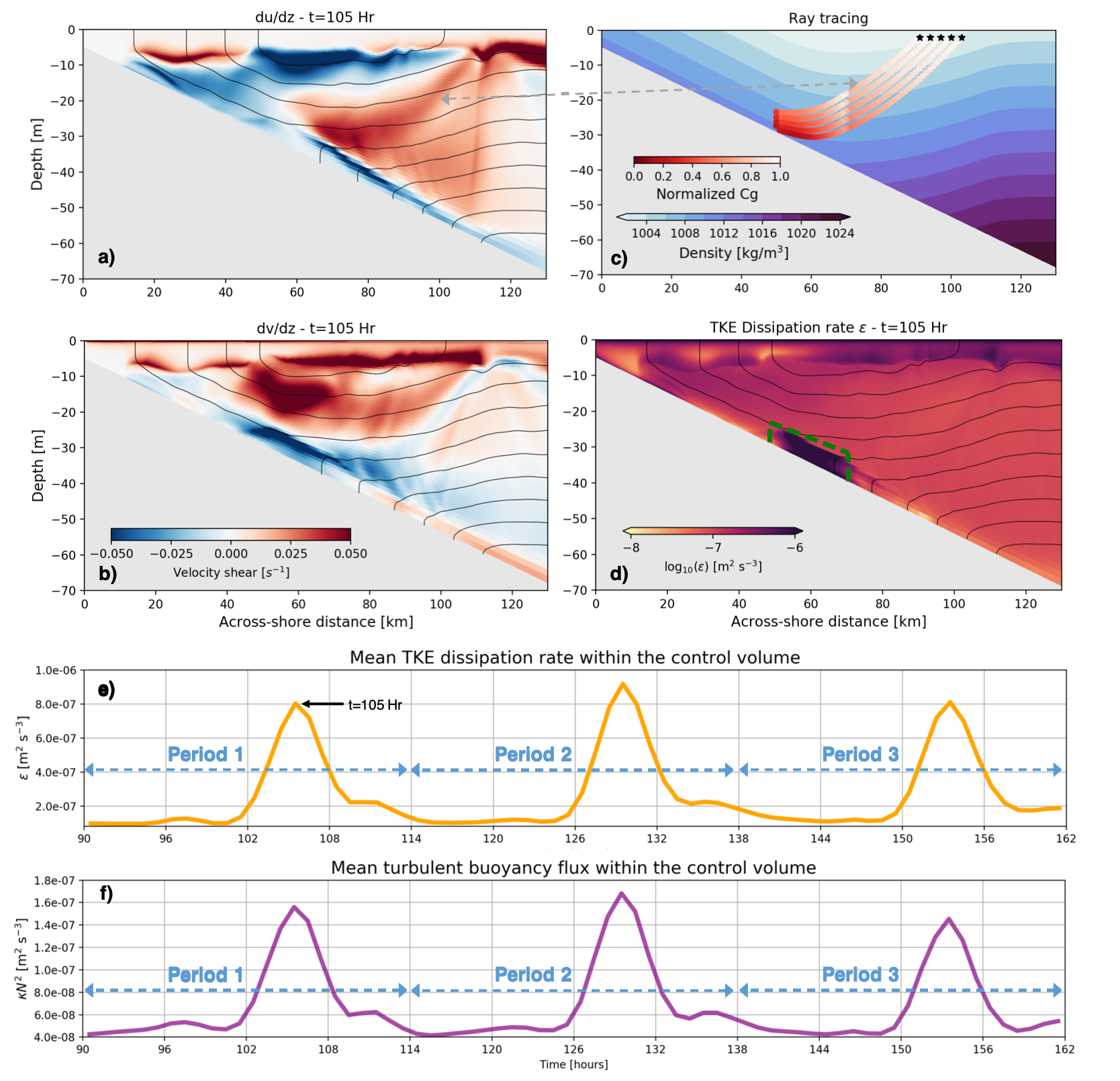}}
\caption{(a and b) Across-shore sections of $\frac{du}{dz}$ and $\frac{dv}{dz}$ from the base run. (c) Ray-tracing solution based on the initial conditions of the base run; the rays are colored by the group velocity (normalized by its maximum value on the ray) and the initial locations of the rays are denoted by the black stars. (d) Across-slope section of TKE dissipation rate $\epsilon$ and the control volume used in the energy budget (green dashed box). Time series of the mean dissipation rate (e) and turbulent buoyancy flux $\kappa N^2$ (f) in the control volume. Three inertial periods are shown in (e) and (f), and the vertical sections are made at t=105 Hr.}
\label{fig4}
\end{figure}

Under the resonant wind forcing, NIWs start to develop in the first few inertial periods and then enhanced bottom mixing follows. Snapshots of the vertical shear after four inertial periods reveal the presence of shear bands (Fig.~\ref{fig4}a and Fig.~\ref{fig4}b). The orientation of the shear bands suggests that the NIWs are generated at the offshore front, where the gradient in relative vorticity is largest. This is consistent with the theoretical prediction that the horizontal wavelength of NIWs shrinks in regions with strong vorticity gradients so that the waves can propagate vertically \citep{young1997propagation, asselin2020penetration}. The slantwise shear bands imply that the NIWs are propagating vertically and bend upwards when approaching the bottom. Furthermore, mixing is enhanced within the bottom critical layer, which corresponds to the area where the waves are focused (Fig.~\ref{fig4}d). This idealized simulation qualitatively reproduces the phenomena found in the realistic simulation (Fig.~\ref{fig1}).

To understand the pattern of wave propagation suggested by the shear bands, ray-tracing is conducted by applying the Wentzel-Kramers-Brillouin (WKB) approximation. The WKB approximation is only valid when the background flow field does not significantly change over the scales of the waves. We will accept this approximation a priori and then validate it below by demonstrating a consistency with an energetics analysis. The initial fields of density and velocity (fig.~\ref{fig3}) are used for the background flow in the ray-tracing calculation. The details of the calculation are described in Appendix B. Rays are initiated at $z=-2m$ at the offshore end of the front with 3 km spacing. The ray paths have a similar shape to the shear bands and indicate that wave energy is radiated downwards from the surface. As the waves enter the anomalously low-frequency regime (marked in the lower panel of fig.~\ref{fig3}), they bend such that the slopes of wave rays are near zero. When the waves approach the critical layer, the waves slow down and eventually get trapped as $|\mathbf{c}_g|\rightarrow 0$ (Fig.~\ref{fig4}c).  Moreover, the rays converge at the location, where the bottom mixing is enhanced, implying that wave trapping within the critical layer might be the mechanism enhancing the bottom mixing.

The mixing in the critical layer exhibits an oscillatory behavior, which is reflected in the temporal variations in the TKE dissipation rate and turbulent buoyancy flux. The TKE dissipation rate $\epsilon$ is diagnosed via the $k-\epsilon$ turbulence closure scheme. The magnitude of the the turbulent buoyancy flux is parameterized as $\kappa N^2$, where $\kappa$ is the turbulent diffusivity also diagnosed from the $k-\epsilon$ closure. The mean values of $\epsilon$ and $\kappa N^2$ are calculated within the control volume (marked by the green box), and the temporal variations of these measures are shown in (Fig.~\ref{fig4}e and \ref{fig4}f). Both $\epsilon$ and $\kappa N^2$ exhibit inertial pulsing, implying that the bottom mixing is enhanced at the inertial frequency. This reproduces the inertial pulsing of $\epsilon$ found in the realistic simulation (Fig.~\ref{fig1}e), and strengthens the link between the bottom enhanced mixing and the NIWs.

\subsection{Energetics}
An energetics analysis is conducted to further support the ray-tracing solution by formulating a kinetic energy equation from the primitive equations for a two-dimensional flow invariant in the $x$-direction:
\begin{equation}
 \begin{aligned}
    \frac{\partial u}{\partial t} + \bm{u} \cdot \nabla u  - f v &=  \frac{\partial}{\partial{z}} (\nu \frac{\partial u}{\partial{z}}),   \\
    \frac{\partial v}{\partial t} + \bm{u} \cdot \nabla v + f u &= -\frac{1}{\rho_0} \frac{\partial{p}}{\partial{y}}  + \frac{\partial}{\partial{z}} (\nu \frac{\partial v}{\partial{z}}),   \\
    -b  &= -\frac{1}{\rho_0} \frac{\partial{p}}{\partial{z}},  \\
    \frac{\partial{u}}{\partial{x}} + \frac{\partial{v}}{\partial{y}} + \frac{\partial{w}}{\partial{z}} &= 0 \ ,    
 \end{aligned}
 \label{eq:prim}
\end{equation}
where $u$ is the along-shore velocity, $v$ is the across-slope velocity, and $w$ is the vertical velocity, $p$ pressure, $b$ is the buoyancy, and the Boussinesq approximation has been assumed. A 24-hour high-pass filter is applied on the pressure and buoyancy to isolate the wave fields $p'$ and $b'$ from the mean fields $\overline{p}$ and $\overline{b}$. These fields are filtered to calculate the wave energy flux convergence (which quantifies the degree of wave trapping), and which involves the wave pressure rather than the total pressure. However, it is not necessary to filter the velocity field. Since $\frac{\partial}{\partial x}=0$ (due to the 2.5D-like model configuration), the $u$-related term drops out from the convergence of the wave energy flux (see Eq.~\ref{eq:KE}), and $v$ and $w$ have weak mean values and are thus dominated by the wave velocities.

A kinetic energy budget can be assessed by formulating the following kinetic energy equation. It is obtained by taking dot product of the momentum equations of Eq.~\ref{eq:prim} with the velocity and applying the continuity equation:
\begin{equation}
 \begin{aligned}
    \frac{\partial KE}{\partial t} &= ADV + WEF + WBF + MEF + MBF + RKE + DKE ,   \\
    KE  &= \frac{1}{2} (u^2+v^2), \\
    ADV   &= - \vec{u} \cdot \nabla KE , \\
    WEF &= - \nabla \ \cdot (\frac{p'}{\rho_0} \vec{u}) = - \frac{\partial}{\partial y} (\frac{p'}{\rho_0}v) - \frac{\partial}{\partial z} (\frac{p'}{\rho_0}w), \\
    WBF &= wb', \\
    MEF &= - \nabla \ \cdot (\frac{\bar{p}}{\rho_0} \vec{u}), \\
    MBF &= w\bar{b}, \\
    RKE &=  \frac{\partial}{\partial z} (\nu \frac{\partial KE}{\partial z}), \\
    DKE  &= -\nu (u_z^2 +v_z^2)  . \\
 \end{aligned}
 \label{eq:KE}
\end{equation}
$ADV$ is the advection of kinetic energy. $WEF$ is the convergence of wave energy flux. $WBF$ is the wave buoyancy flux representing the energy transfer between wave kinetic and potential energy. $MEF$ and $MBF$ are the mean energy flux convergence and buoyancy flux, and go to zero when averaged over times greater than a wave period. $RKE$ represents the redistribution of kinetic energy by turbulence. $DKE$ is the dissipation of kinetic energy, representing the loss of mean and wave kinetic energy to turbulence. 

\begin{figure}[t]
\centerline{\includegraphics[width=\textwidth]{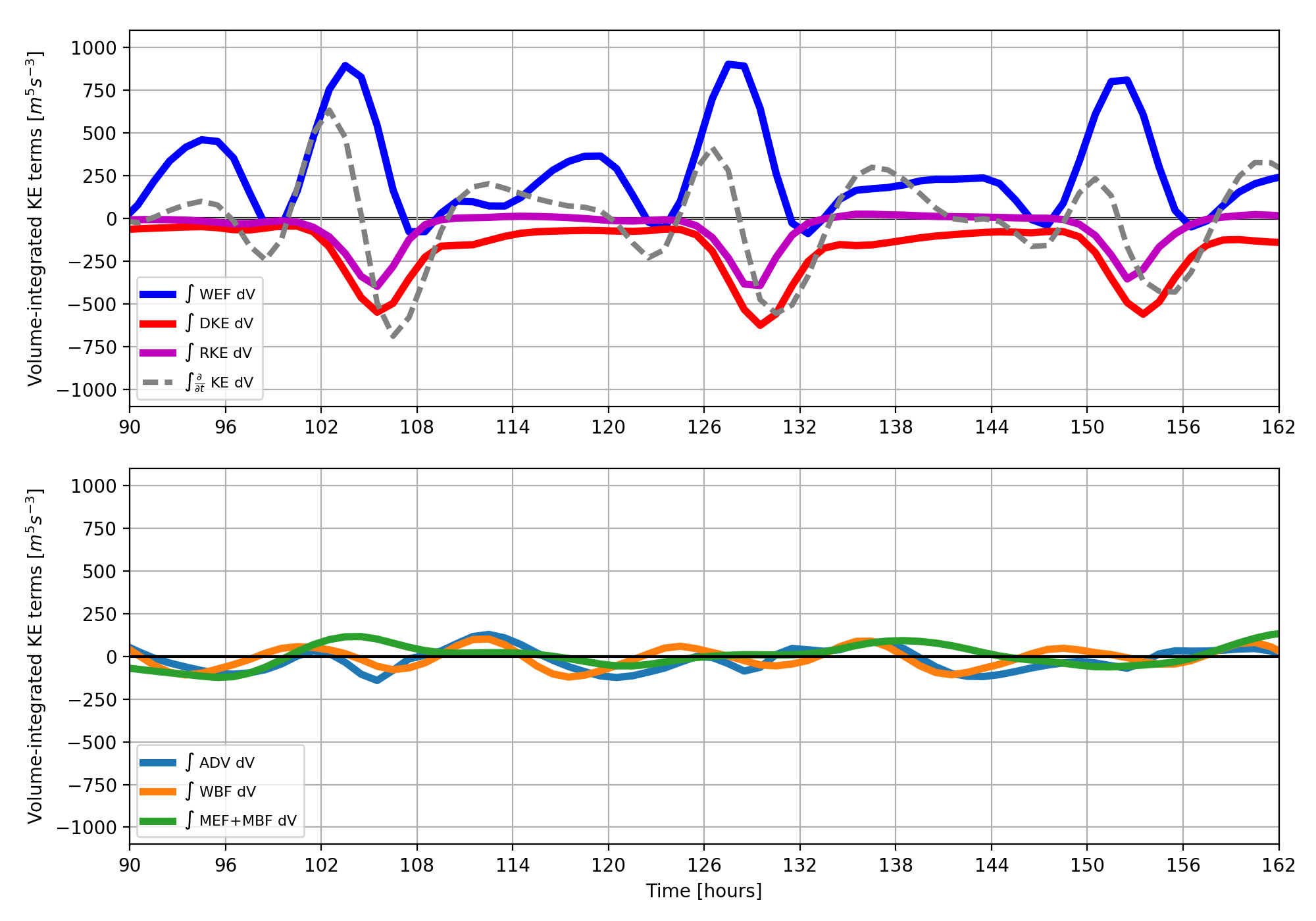}}
\caption{Time series of the control-volume integrated terms of Equation (\ref{eq:KE}). The dominant terms are shown in the upper panel and the less significant terms in the lower panel.}
\label{fig5}
\end{figure}

The energetics of the waves in the idealized simulation are analyzed, using Eq.~\ref{eq:KE}, within the control volume marked by the green box in Fig.~\ref{fig4}d, where the mixing is enhanced. Each term in Eq.~\ref{eq:KE} is integrated over the control volume to obtain time series (Fig.~\ref{fig5}). $\int\ WEF\  dV$ represents the wave energy flux coming into (positive) or going out of (negative) the control volume. Since the turbulent viscosity parameterizes turbulent momentum fluxes, it follows that $DKE$ is related to the shear production terms in the TKE equation, and thus $\int\ DKE\ dV$ represents the portion of the Reynolds-averaged kinetic energy transferred to the TKE, which would further go into the turbulent buoyancy flux $\kappa N^2$ and the TKE dissipation $\epsilon$ (see Section 5.b for more details). $\int\ RKE\ dV$ quantifies the energy removal out of the control volume by the bottom stress, since it can be rewritten as $\int_{A_b}\ \bm{u} \cdot \vec{\tau_b}\ dS$ (where $\vec{\tau_b}$ is the bottom stress and $A_b$ is the bottom boundary) if the stress at the top boundary of the control volume is negligible. A clear inertial pulsing is found in the time series of $\int\ DKE\ dV$, which follows the inertial pulsing of $\epsilon$ and $\kappa N^2$ (Fig.~\ref{fig4}e and \ref{fig4}f) and hence the inertially enhanced mixing. The time series of $\int\ WEF\  dV$ exhibits peaks that lead $\int\ DKE\ dV$ by 1 hour, implying that the convergence of wave energy causes the enhanced bottom mixing. The inertial pulsing of $\int\ RKE\ dV$ suggests that the bottom stress removes energy when the wave energy flux converges and the flow enhances the turbulence. All the other terms in the KE budget are less significant (lower panel of Fig.~\ref{fig5}). Overall, given the high correlation between $\int\ DKE\ dV$ and $\int\ WEF\  dV$, the process driving the enhanced mixing is wave trapping, consistent with the inference based on ray-tracing.

\subsection{Comparative run}
\begin{figure}[t]
\centerline{\includegraphics[width=\textwidth]{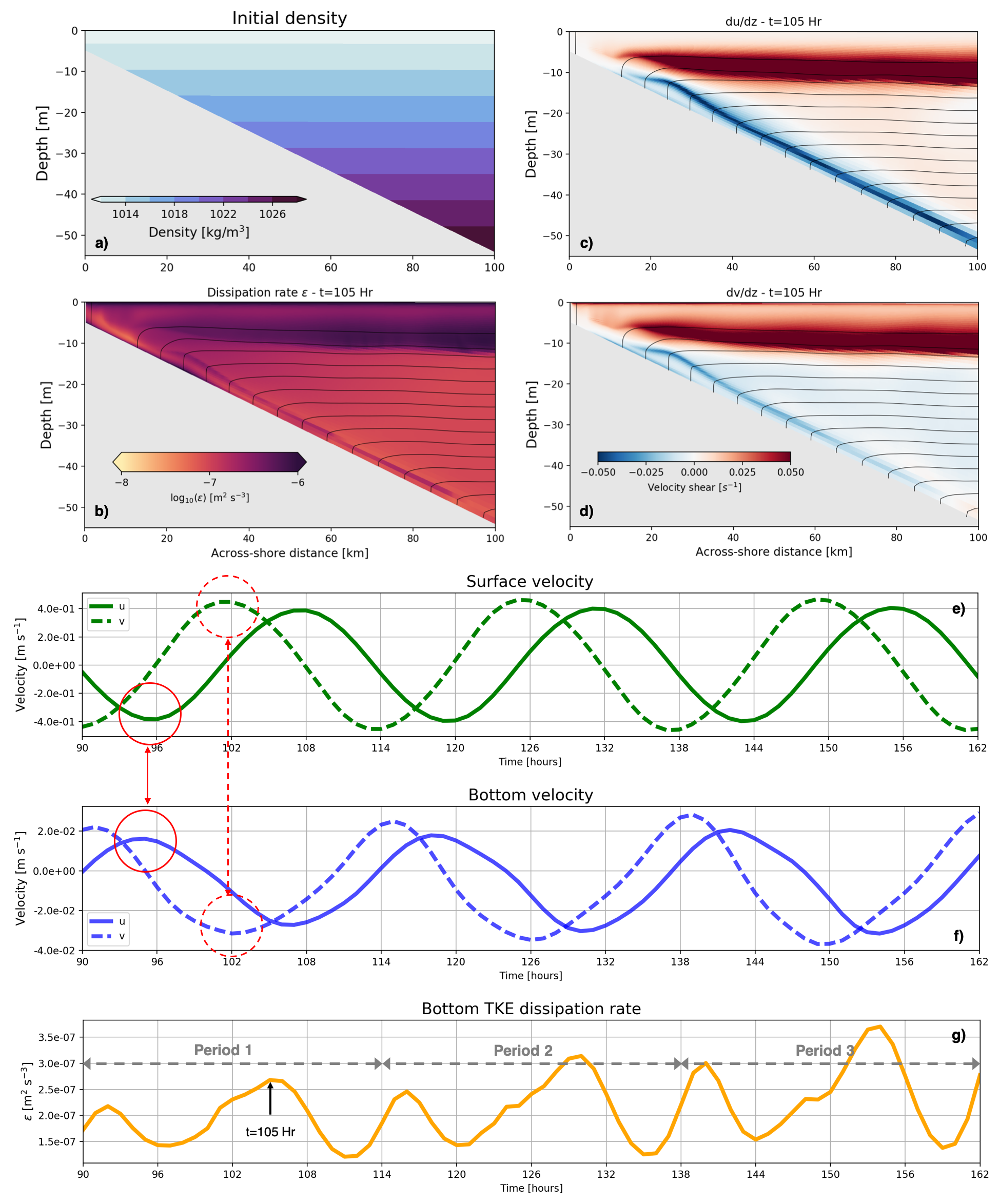}}
\caption{(a) Initial density of the comparative run. (b, c and d) Across-shore sections of $\epsilon$, $\frac{du}{dz}$, and $\frac{dv}{dz}$ at t=105 Hr. (e,f, and g) Time series of the surface velocity, bottom velocity, and bottom TKE dissipation rate; these quantities are averaged within the across-shore distance of 100 km. The maxima and minima of velocity are denoted by red circles. }
\label{figcontrast}
\end{figure}

To demonstrate the effect of coastal fronts on vertically-radiating NIWs, we present a comparative simulation without the eddy-like front to contrast the response of NIWs and bottom mixing to the simulation with the eddy-like front (that is, the base run discussed above). The initial density field is shown in Fig.~\ref{figcontrast}. The difference with the setup of the base run is the absence of the lateral buoyancy gradients, and hence also no background flow, all other parameters are the same (Tab.~\ref{tab1}). 

A "two-layer" response is found to be dominant in the comparative simulation; the surface and bottom velocities are out of phase by nearly 180 degrees (Fig.~\ref{figcontrast}e and \ref{figcontrast}f). Such a "two-layer" structure has been observed in many coastal seas \citep{orlic1987oscillations,van1999strong,knight2002inertial,rippeth2002current} and also some large lakes \citep{malone1968analysis,smith1972temporal}. This response is attributed to the presence of a coastal boundary \citep{davies2002influence}. The presence of the coastal boundary yields a pressure gradient at depth, which drives the inertial current in the lower layer and leads to a 180 degree phase shift with the upper layer \citep{xing2004influence}. Consistent with the observations, the comparative run shows that the first baroclinic mode dominates the response in a coastal system without fronts or currents.

The differences between the comparative and base runs are summarized as follows. First, no clear slantwise shear bands exist in the interior (Fig.~\ref{figcontrast}c and \ref{figcontrast}d), suggesting that there is no significant vertical radiation of NIWs from the surface. Second, the dissipation does not exhibit clear inertial pulsing that has been observed in the base run (Fig.~\ref{figcontrast}g); it pulses semi-inertially as the bottom velocity attains peaks. Lastly, the bottom boundary layer over the shelf is thin (Fig.~\ref{figcontrast}b), and the response of the dissipation is much weaker than that in the base run (Fig.~\ref{figcontrast}e and ~\ref{figcontrast}g). Overall, the comparative simulation indicates that the eddy-like coastal front is essential for the vertical radiation of NIWs and therefore the bottom enhanced mixing.

\subsection{Parameter dependence of wave trapping and mixing in the critical layer} 
The idealized simulations are used to explore the dependence of wave trapping and mixing in the critical layer in the framework of the idealized configuration. There are 8 controlling parameters and they are listed in Tab.~\ref{tab1}. To efficiently explore this parameter space, we fix the external parameters (i.e., the background rotation $f$, wind forcing $\vec{\tau}$, and bottom slope $\alpha$) as well as the dimension of the near-shore front (i.e., the length scales of the critical layer and transition zone, $L_C$ and $L_T$), but vary the parameters associated with the offshore front (i.e., the frontal width $L$, relative vorticity $\zeta_0$, and stratification $N^2$). In other words, we fix the properties of the critical layer but vary the parameters that influence the propagation of waves towards the critical layer. By varying these parameters we can quantify the sensitivity of the dissipation and mixing to the degree of wave trapping. A total of 18 simulations (including the base run) were performed and are listed in the second row of Tab.~\ref{tab1}. 

In concert with the ROMS simulations, ray-tracing is conducted for each run. Rays are initiated at $z=-2\ m$ within the offshore front across the width $L$, separated by a spacing of $dy=1\ km$, and traced according to the procedure described in Appendix B. Rays either reach the critical layer or hit the bottom and reflect offshore. To quantify the wave trapping in the critical layer from the ray-tracing solutions, we define the trapping ratio $\gamma$, the ratio between the number of the rays reaching the critical layer and the total number of the rays. A higher value of $\gamma$ indicates a larger portion of wave energy that reaches the trapping zone and hence represents a highly trapped scenario. The trapping ratio is a metric that concisely captures the parameter dependence of wave trapping in the critical layer.

\begin{figure}[t]
\centerline{\includegraphics[width=1.1\textwidth]{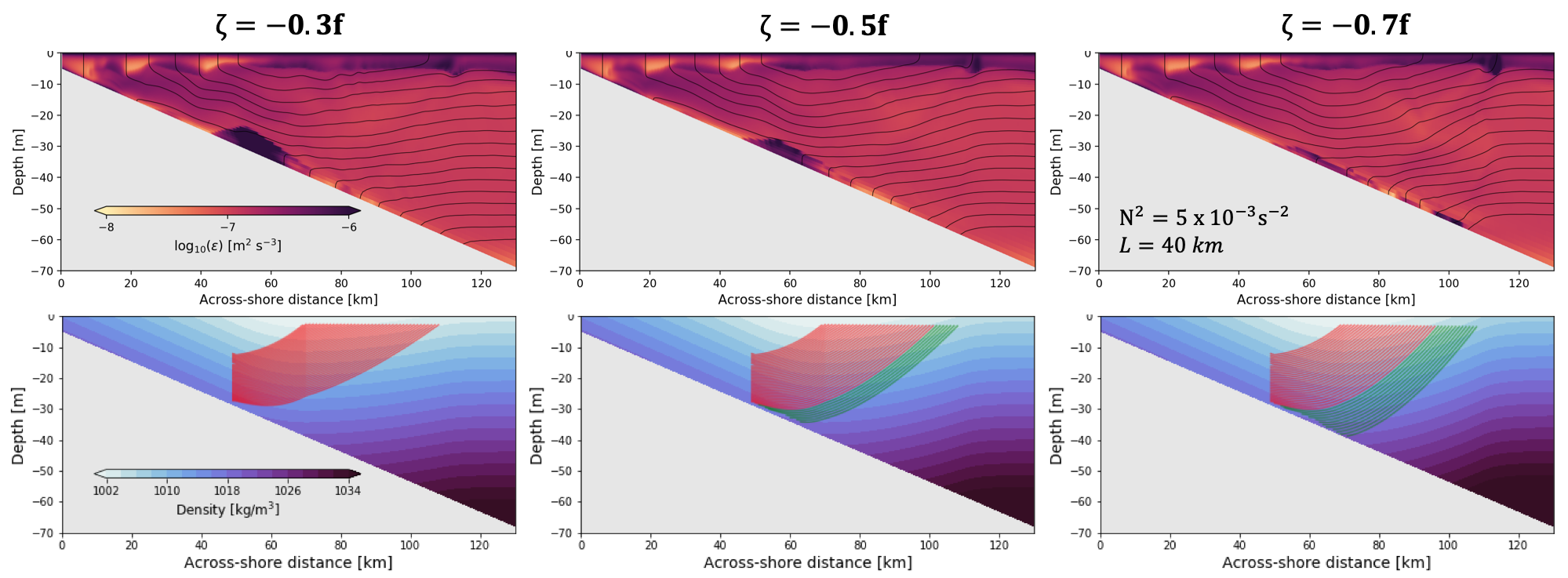}}
\caption{(upper) Across-slope sections of TKE dissipation rate $\epsilon$ from the ensemble runs with a stratification of $N^2=5 \times 10 ^{-3} s^{-2}$, offshore frontal width of $L=40\ km$, and varying surface vorticity values ($\zeta_0=-0.3f$, $-0.5f$, and $-0.7f$). The sections are made at t=105 Hr. (lower) Rays for each run; red denotes the rays that enter the critical layer, and green denotes the rays that reflect off the bottom and away from the critical layer.}
\label{fig6}
\end{figure}

Relative vorticity modifies the minimum frequency $\omega_{min}$, such that stronger anti-cyclonic vorticity allows NIWs to propagate more vertically, e.g. (\ref{ray_slope}). A subset of the ensemble simulations run with different values of the vorticity $\zeta_0$ but with a fixed frontal width of $L=40 \ km$, and stratification of $N^2 = 5 \times 10 ^{-3} s^{-2}$ illustrates this physics (Fig.~\ref{fig6}). The stronger the anti-cyclonic vorticity, the more steep the rays are, and they miss the critical layer. For instance, in the case with $\zeta_0=-0.7f$, there are fewer rays reaching the critical layer (denoted by red) and more rays reflect offshore (denoted by green) than in the other two runs with weaker vorticity. Correspondingly, the case with $\zeta_0=-0.7f$ has the lowest TKE dissipation rate, suggesting that reduced wave trapping leads to weaker mixing.

\begin{figure}[t]
\centerline{\includegraphics[width=1.1\textwidth]{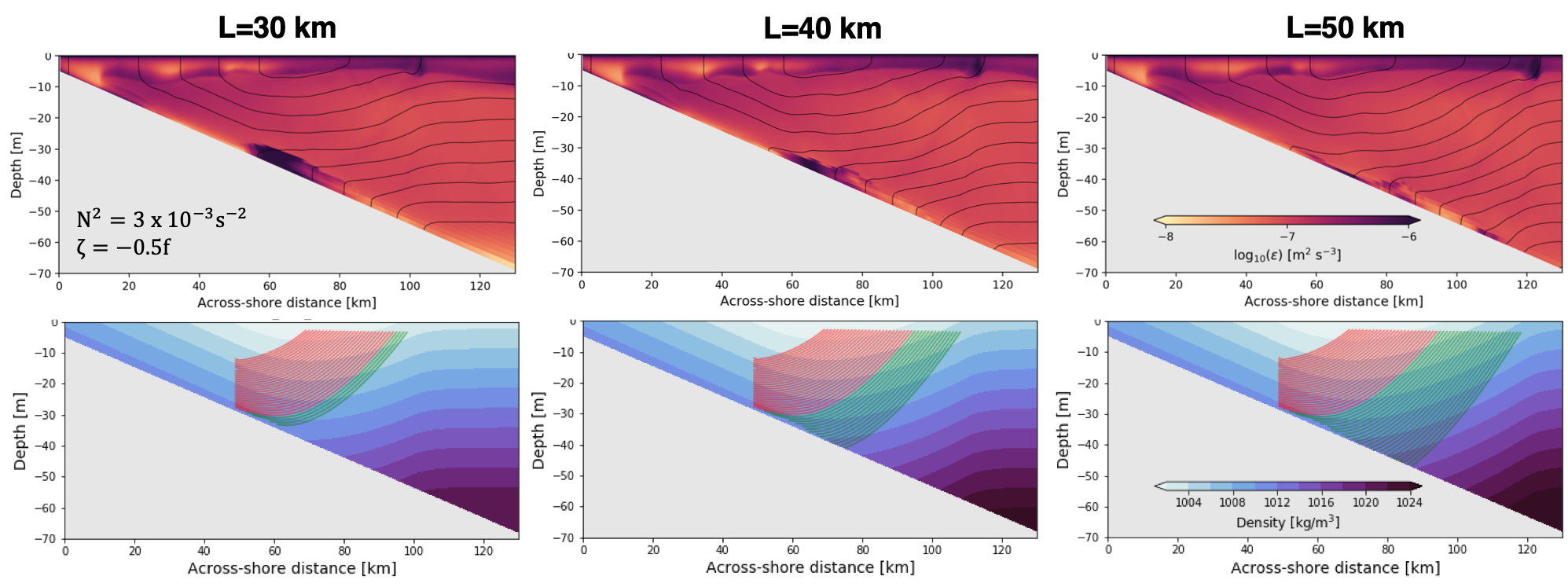}}
\caption{Same as Fig.~\ref{fig6} but for the ensemble runs with a stratification of $N^2=3 \times 10 ^{-3} s^{-2}$, offshore surface vorticity of $\zeta_0=-0.5f$, and varying frontal widths ($L=30\ km$, $40\ km$, and $50\ km$).}
\label{fig7}
\end{figure}

The dimension of the offshore front also modulates the wave trapping. This is illustrated in Fig.~\ref{fig7} for a group of ensemble runs with various frontal widths $L$ but with fixed relative vorticity (i.e. $\zeta_0=-0.5f$) and stratification ($N^2 = 3 \times 10 ^{-3} s^{-2}$). Noting that the difference in vorticity across the jet is the same for all three cases, the propagation of NIWs only depends on the geometry of the offshore front. The ray-tracing solutions shown in Fig.~\ref{fig7} demonstrate that the wider fronts have fewer rays that reach the critical layer and get trapped. This is because wider fronts move the rays away from the critical layer. Consquently, the run with $L=50\ km$ has the weakest TKE dissipation rate and mixing.

\begin{figure}[t]
\centerline{\includegraphics[width=1.1\textwidth]{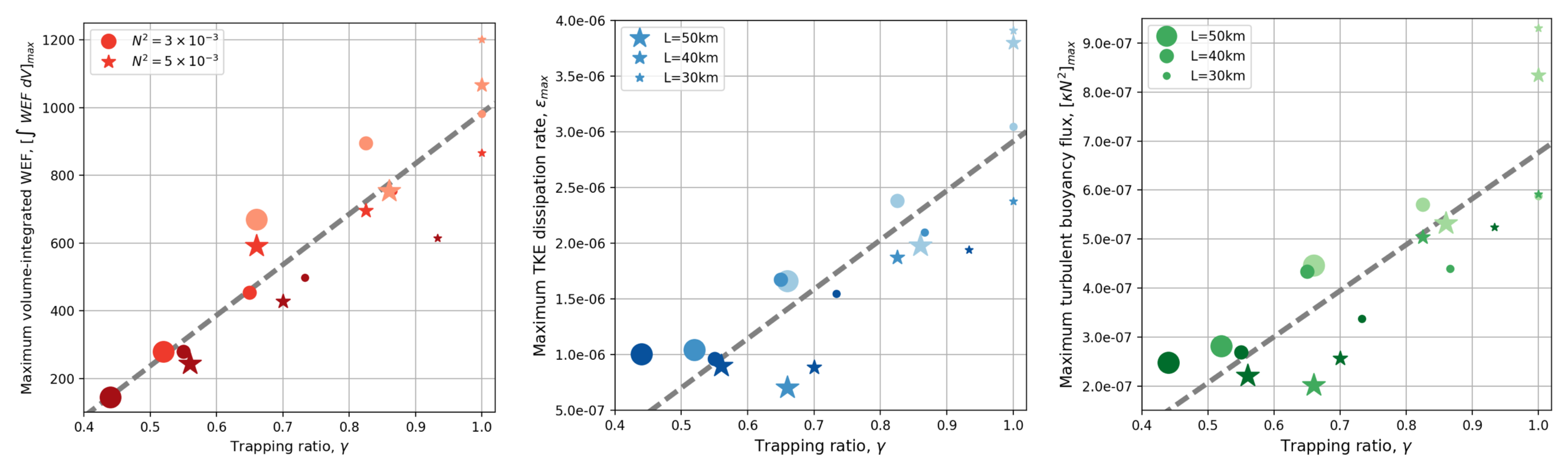}}
\caption{Maximum volume-integrated convergence of wave energy flux $[\int\ WEF\ dV]_{max}$ (left), TKE dissipation rate $\epsilon_{max}$ (middle), and turbulent buoyancy flux $[\kappa N^2]_{max}$ (right) plotted against the trapping ratio $\gamma$. The maxima are calculated over Inertial Period 1. The runs with $N^2=3\times 10^{-3} s^{-2}$ are marked by circles, and the runs with $N^2=5\times 10^{-3} s^{-2}$ are marked by stars. A larger marker size represents a larger offshore frontal width, and a darker color represents stronger anticyclonic vorticity. The gray dashed lines indicate the linear regressions. }
\label{fig8}
\end{figure}

Finally, we quantify the relation between wave trapping and mixing using the trapping ratio $\gamma$. First, to test the skill of the parameter $\gamma$ in predicting the degree of wave trapping, we calculate the maximum (in time, over one inertial period) volume-integrated WEF $[\int\ WEF\ dV]_{max}$ for each run and compare this quantity to $\gamma$. Recall that a large value of $[\int\ WEF\ dV]_{max}$ corresponds to strong wave trapping. The trapping ratio $\gamma$ and $[\int\ WEF\ dV]_{max}$ are highly correlated ($r=0.91$ and $p=1.15\times10^{-7}$; left panel of Fig.~\ref{fig8}), suggesting that $\gamma$ is a skillful predictor for wave trapping. Next, $\gamma$ is compared to the maximum (over one inertial period and within the control volume) TKE dissipation rate $\epsilon_{max}$ and the maximum turbulent buoyancy flux $[\kappa N^2]_{max}$ (Fig.~\ref{fig8}). The correlations between these quantities are also robust: $r=0.85$ and $p=9.26\times10^{-6}$ for $\epsilon_{max}$ and $r=0.84$ and $p=1.29\times10^{-5}$ for $[\kappa N^2]_{max}$. This indicates that strong trapping of wave rays leads to high turbulence dissipation and mixing. Overall, the ensemble runs further support the conclusion, that the enhanced bottom mixing is caused by wave trapping in the bottom critical layer.

\section{Discussion}\label{discussion_section}
\subsection{Enhanced diapycnal transport in the critical layer}
\begin{figure}[t]
\centerline{\includegraphics[width=\textwidth]{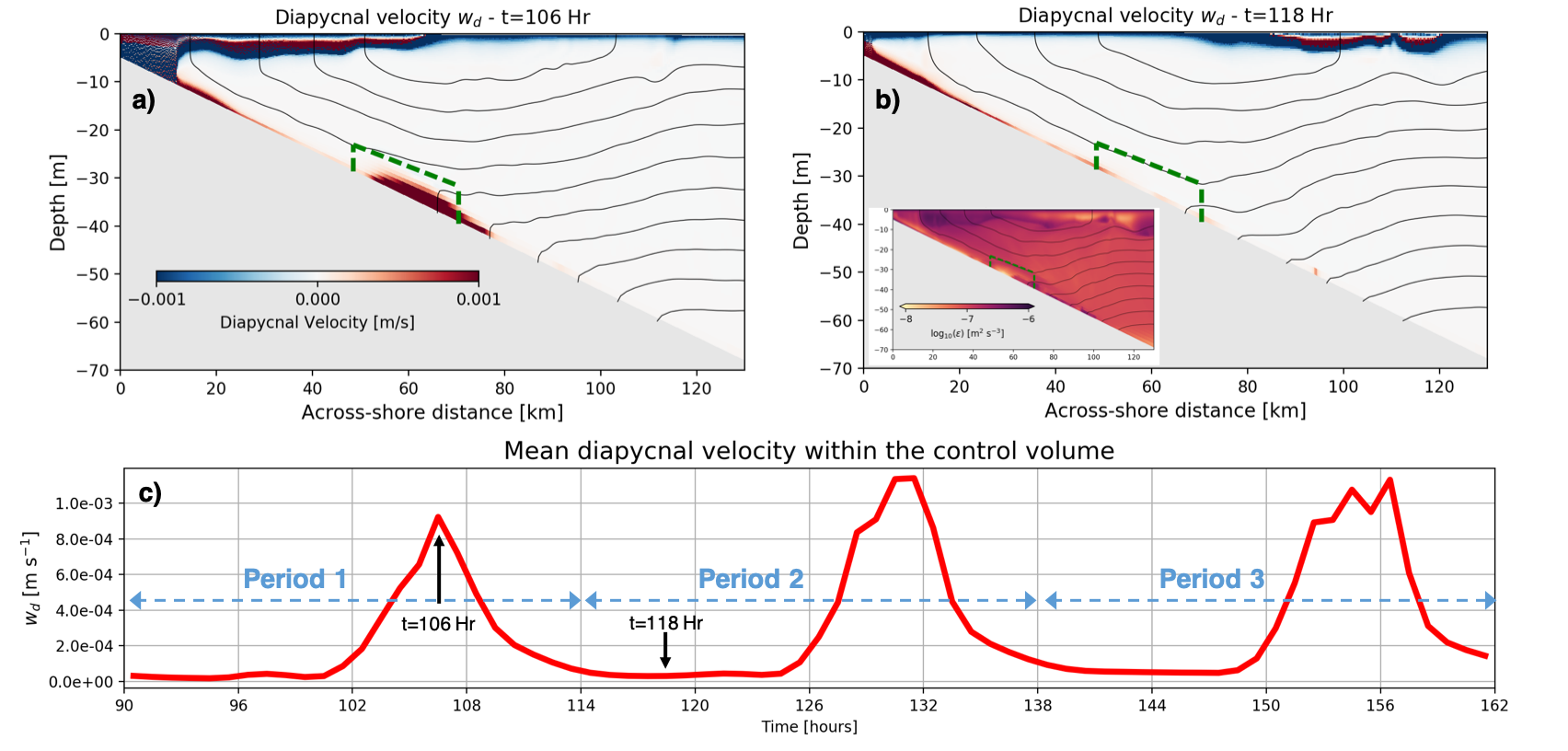}}
\caption{Across-slope sections of diapycnal velocity $w_d$ from the base run at t=106 Hr (a) and t=118 Hr (b). The subpanel in (b) is the across-slope section of $\epsilon$ at t=118 Hr. (c) Time series of the diapycnal velocity averaged in the control volume (green dashed box). Three inertial periods are shown.}
\label{fig_DiaVel}
\end{figure}

\begin{figure}[t]
\centerline{\includegraphics[width=0.5\textwidth]{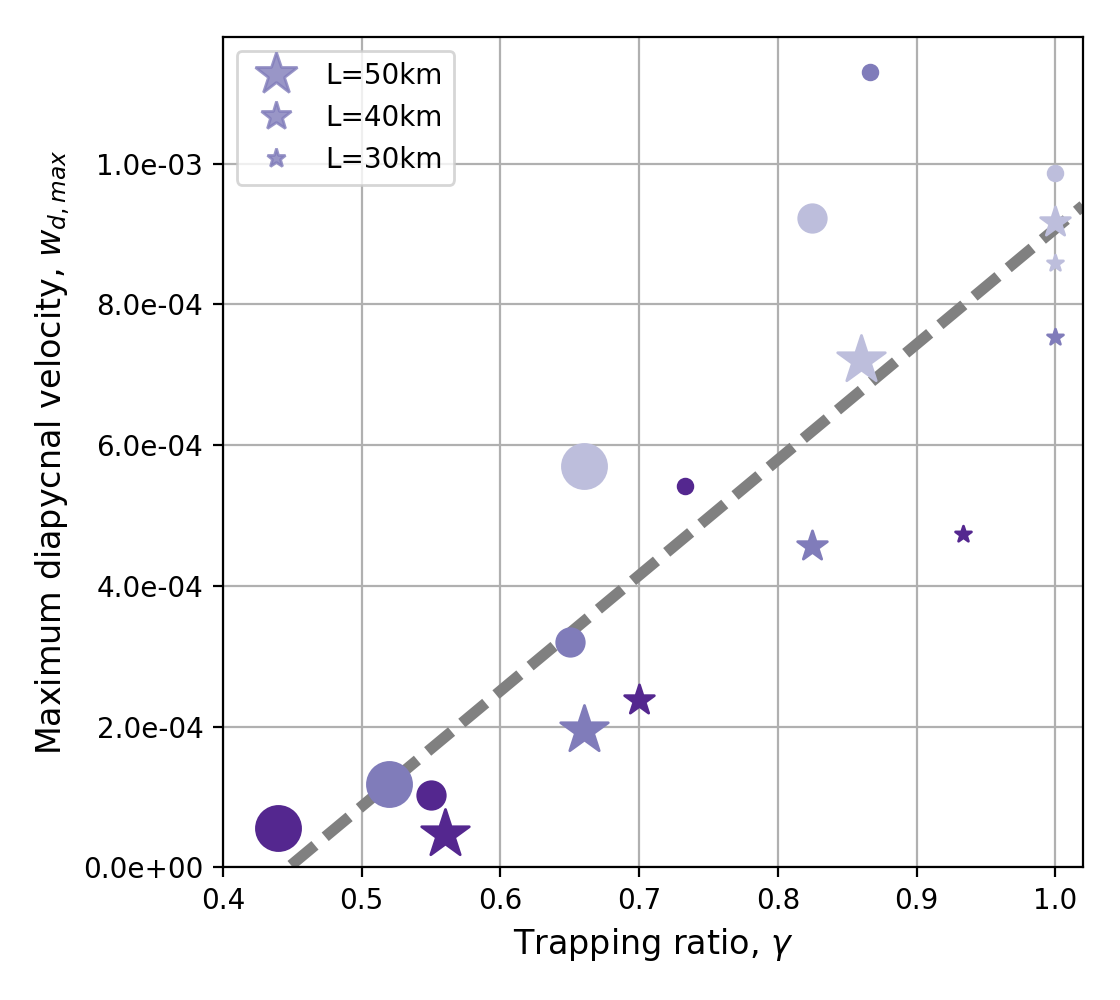}}
\caption{Maximum volume-averaged diapycnal velocity $w_{d,max}$ plotted against the trapping ratio $\gamma$. The maximum is calculated in Inertial Period 1. The size, shape, and shading of the markers are the same as in Fig.~\ref{fig8}. The gray dashed line indicates the linear regression.}
\label{fig_gamma_wd}
\end{figure}

The amplification of NIWs by wave trapping, which elevates mixing, also enhances diapycnal transport. To examine the link between the diapycnal transport and wave trapping in the slantwise critical layer, the diapycnal velocity,
\begin{equation}
    w_d = \frac{\frac{\partial}{\partial z}(\kappa N^2)}{N^2},
 \label{eq:wd1}
\end{equation}
is diagnosed across the idealized simulations listed in Tab.~\ref{tab1}. The influence of lateral density gradients is neglected in Eq.~(\ref{eq:wd1}), which is justified since the isopycnal slopes are very small  \citep{bennett1986relationship}. The diagnostics shows that the diapycnal velocity is upward in the critical layer and enhanced when the bottom mixing is elevated (Fig.~\ref{fig_DiaVel}a). Also, the diapycnal velocity decays as the mixing weakens (Fig.~\ref{fig_DiaVel}b), so that the time series of the diapycnal velocity has a similar inertial pulsing as the bottom mixing (Fig.~\ref{fig_DiaVel}c). Note that the amplitude of the diapycnal velocity can reach $O(10^{-3})$ m/s, which is strong and comparable to the entrainment velocity near the surface induced by wind-driven turbulence. Furthermore, the maximum volume-averaged diapycnal velocity $w_{d,max}$ (i.e. the maximum over one inertial period) is robustly correlated with the trapping ratio $\gamma$, with $r=0.85$ and $p=7.93\times10^{-6}$ (Fig.~\ref{fig_gamma_wd}). This further strengthens the link between the enhancement of the diapycnal velocity and the wave trapping mechanism. 

\begin{figure}[t]
\centerline{\includegraphics[width=\textwidth]{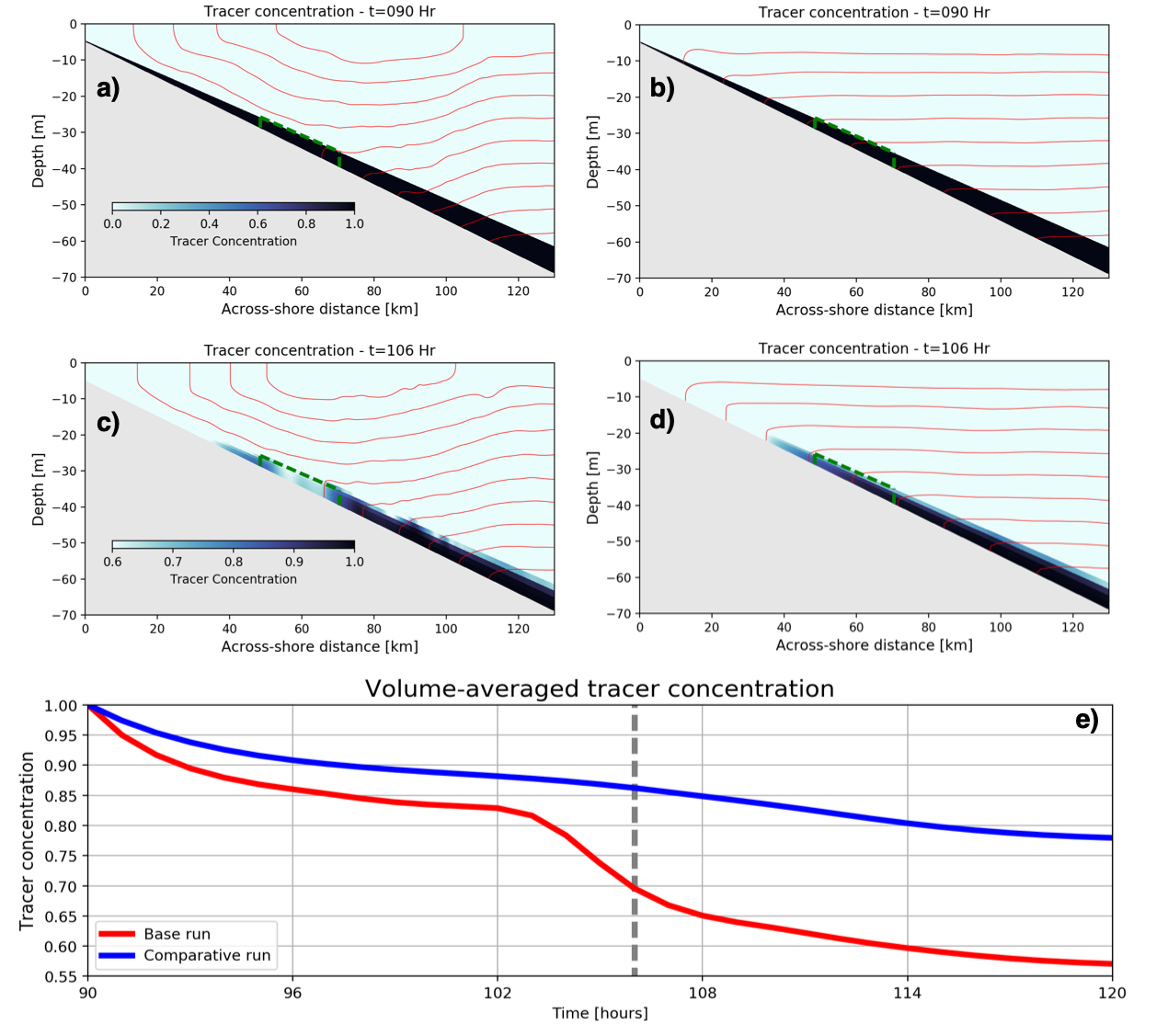}}
\caption{Initial tracer field in the base run (a) and the comparative run (b). Tracer field at t=106 Hr in the base run (c) and the comparative run (d). (e) Time series of the volume-averaged tracer concentration in the base run and the comparative run. The volume used in the calculation is marked by the green dashed box. The gray dashed line in (e) denotes t=106 Hr.}
\label{fig_bottom_tracer}
\end{figure}

Another way to quantify the diapycnal transport is to track the diapycnal movement of a passive tracer. To this end, a passive tracer was released in both the base run and the comparative run to contrast the bottom diapycnal transport in simulations with and without wave trapping. The tracer is initialized in the first four sigma layers above the bottom with a concentration equal to one (Fig.~\ref{fig_bottom_tracer}a and ~\ref{fig_bottom_tracer}b). The concentration outside of this layer is set to zero. The tracer is released at t=90 Hr and monitored for three inertial periods. In terms of the spatial distribution, at t=106 Hr (the time of the peak $w_d$), the base run shows a significant reduction of the tracer at the location where the diapycnal velocity is enhanced and the waves are trapped (Fig.~\ref{fig_bottom_tracer}c). In terms of temporal variability, in the region with enhanced $w_d$, the variation of the tracer concentration suggests that the tracer is transported out of the bottom layer during the period (from 103 Hr to 109 Hr) when the diapycnal velocity is enhanced (Fig.~\ref{fig_bottom_tracer}e). In contrast, the comparative run does not show such a significant reduction in the tracer concentration near the bottom (Fig.~\ref{fig_bottom_tracer}d,e).

The tracer distribution in the density space is used as a metric for tracking the diapycnal tracer transport. The metric is calculated as follows. Given a volume where the density is less than a certain density $\rho$, the tracer content $M$ in this volume is equal to the volume integral of the tracer concentration $C$:
\begin{equation}
		M(\rho,t)=\int\int\int_{\rho'<\rho} C(x,y,z,t)\ dxdydz \  .
 \label{eq:Tmass}
\end{equation}
Then, the tracer distribution function is defined as $\frac{\partial M(\rho,t)}{\partial \rho}$ such that the tracer content within a density class can be obtained by integrating the distribution function in the density space as $\int_{\rho_1}^{\rho_2} \frac{\partial M}{\partial \rho} d\rho$. In other words, $\frac{\partial M}{\partial \rho}$ indicates the instantaneous distribution of the tracer in the density space and any diapycnal tracer transport should be reflected by the rate of change of $\frac{\partial M}{\partial \rho}$.

\begin{figure}[t]
\centerline{\includegraphics[width=0.8\textwidth]{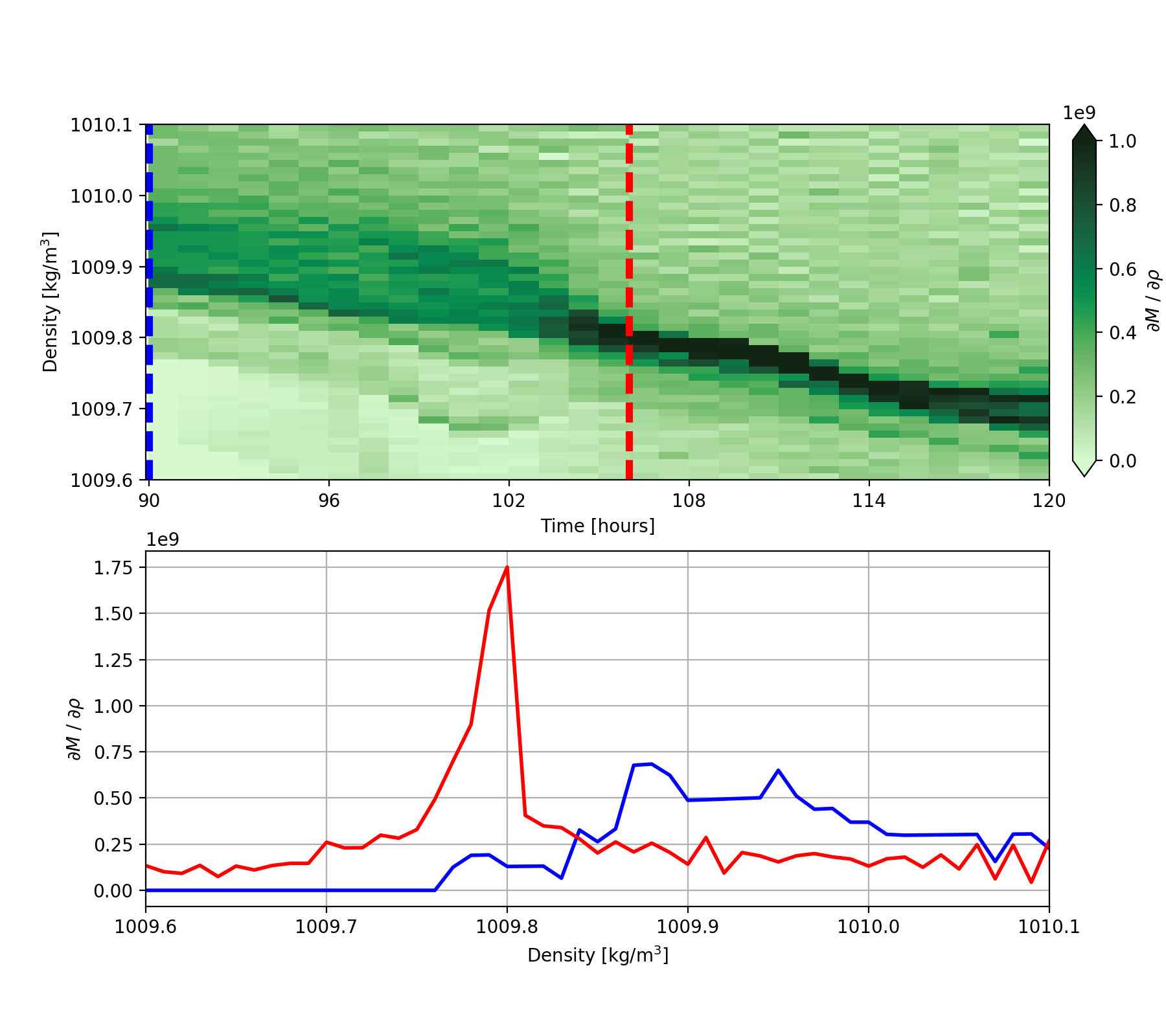}}
\caption{(upper) Hovmöller diagram of the tracer distribution function $\frac{\partial M}{\partial \rho}$ in density space from the base run. $\frac{\partial M}{\partial \rho}$ is calculated between 50 km to 70 km in the cross-shore direction. (lower) $\frac{\partial M}{\partial \rho}$ at t=90 Hr (blue line) and t=106 Hr (red line). The times when $\frac{\partial M}{\partial \rho}$ is evaluated are indicated in the upper panel.}
\label{fig_Rho}
\end{figure}

The distribution function calculated from the base run indicates that the tracer migrates to lighter density classes with time (see the upper panel of Fig.~\ref{fig_Rho}). When the diapycnal velocity is largest (at t=106 Hr), there is a convergence of the tracer towards a narrow density class. This highlights the role of the diapycnal velocity in transporting the tracer. Also, the convergence can be seen by contrasting $\frac{\partial M}{\partial \rho}$ at the time when $w_d$ is maximum with the one at the initial time (see the lower panel of Fig.~\ref{fig_Rho}). Furthermore, the convergence of the tracer persists with time, confirming that the enhanced diapycnal velocity does effectively transport the tracer across isopycnals.

The enhancement of the diapycnal transport by wave trapping has implications for coastal biogeochemistry and ecosystems. In coastal zones, freshwater from rivers strengthens the stratification and can suppresses the ventilation of bottom waters. This combined with phytoplankton blooms fueled by nutrients in the freshwater can lead to bottom hypoxia and the formation of "dead zones" \citep{bianchi2010science}. One region where bottom hypoxia often occurs is the Texas-Lousiana shelf where the development of the hypoxia is heavily modulated by near-inertial motions \citep{xomchuksmall}. We have demonstrated that NIW trapping within critical layers is potentially active, suggesting that mixing of the stratified bottom waters by this process could potentially ventilate these oxygen poor waters. In fact, intrusions of hypoxic waters emanating from slantwise stratified layers near the bottom have been observed on the shelf during the MCH survey suggesting active mixing in these layers \citep{zhang15}.

\subsection{The inertial pulsing of mixing}
A prominent feature of the mixing in the critical layer is its inertial pulsing, which is evident in the variations of the TKE dissipation rate, $\epsilon$, and the turbulent buoyancy flux, $B=\kappa N^2$, in the realistic and idealized simulations (Fig.~\ref{fig1} and \ref{fig4}). In the $k-\epsilon$ turbulent closure model used in the simulations, the direct energy supply for $\epsilon$ and $B$ is the turbulent shear production $P$, which is parameterized as
\begin{equation}
		P = \nu [(\frac{\partial{u}}{\partial{z}})^2+(\frac{\partial{v}}{\partial{z}})^2] \  
 \label{eq:P}
\end{equation}
 \citep{warner2005performance}. Noticing that $DKE=-\nu [(\frac{\partial{u}}{\partial{z}})^2+(\frac{\partial{v}}{\partial{z}})^2]$ (see Eq.~(\ref{eq:KE})), it follows that $P=-DKE$, and hence the terms that balance DKE in the Reynolds-averaged kinetic energy equation (\ref{eq:KE}) are also the source of energy for $\epsilon$ and $B$. The analysis of the energetics (Fig.~\ref{fig5}) shows clear inertial pulsing in $DKE$ in response to $WEF$, indicating that the inertial variations in mixing are ultimately caused by the convergence of the wave energy flux.  

\begin{figure}[t]
\centerline{\includegraphics[width=1.0\textwidth]{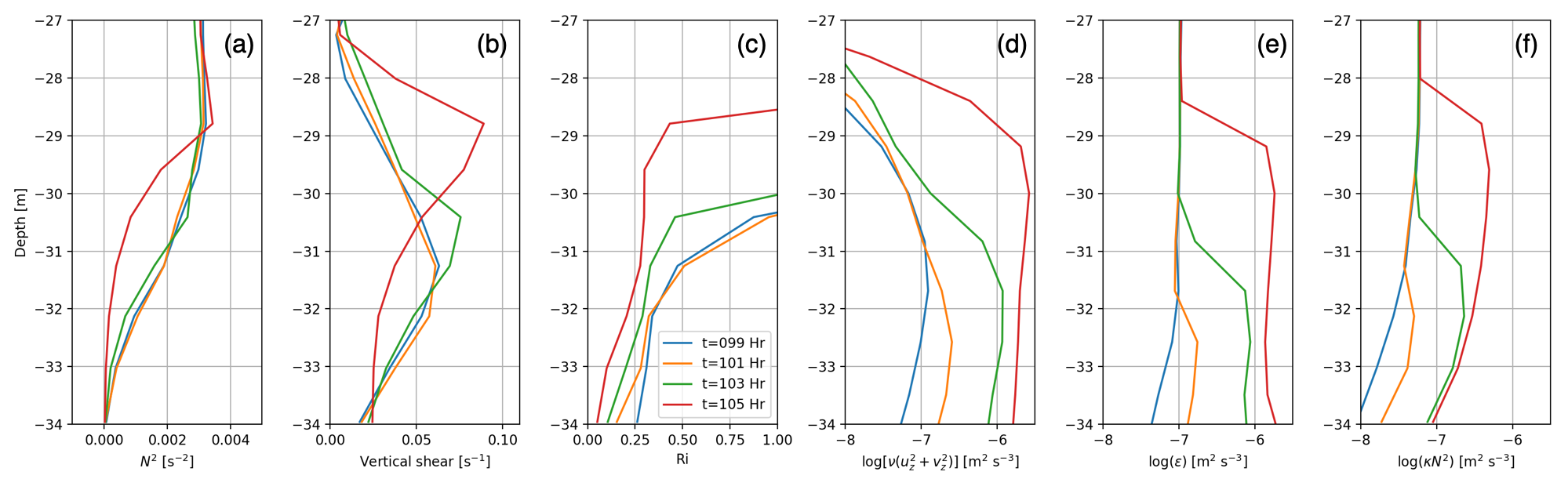}}
\caption{Vertical profiles at the center of the control volume ($y=60\ km$) from the base run. The quantities from left to right are the stratification, vertical shear ($\sqrt{(\frac{du}{dz})^2+(\frac{dv}{dz})^2}$), Richardson number, turbulent shear production, TKE dissipation rate, and turbulent buoyancy flux. The profiles are made during the onset of the enhanced mixing (e.g. between 99-105 hours).}
\label{fig_onset}
\end{figure}

 The influence of $WEF$ in shaping the evolution of the vertical structure of the shear, stratification, dissipation, and mixing is shown in (Fig.\ref{fig_onset}). As the wave energy flux converges (e.g.  Fig.~\ref{fig5} between hours 99-105), the vertical shear and $DKE$ increases (Fig.\ref{fig_onset}b,d). Consequently, $\epsilon$ and $\kappa N^2$ are enhanced via the $k-\epsilon$ model (Fig.\ref{fig_onset}e and f) and the stratification is reduced (Fig.\ref{fig_onset}a). At this stage, the total Richardson number, $Ri$, decreases to around 0.25, indicating that the criterion for shear instabilities is crossed (Fig.\ref{fig_onset}c).

It might seem counterintuitive that the wave energy flux pulses inertially rather than semi-inertially. Based on the polarization relation for linear NIWs, the wave pressure and across-front velocity are in phase so that the wave energy flux should exhibit a semi-inertial response \citep{whitt2013near}. This theoretical prediction breaks down in the simulations because the waves are of finite amplitude and hence their dynamics is nonlinear.  

\begin{figure}[t]
\centerline{\includegraphics[width=0.85\textwidth]{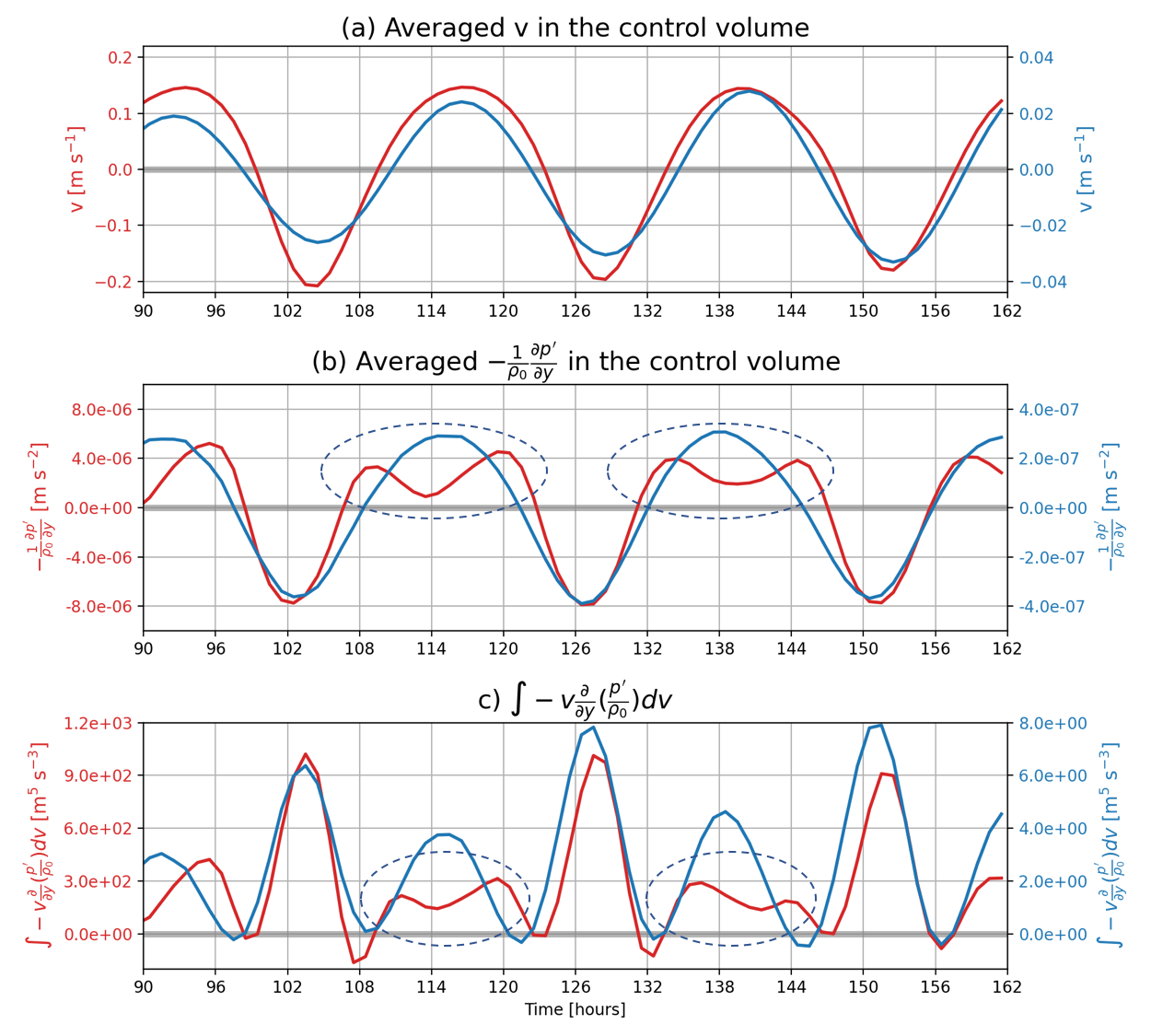}}
\caption{Comparison of $v$ (a), $-\frac{1}{\rho_0}\frac{\partial{p'}}{\partial{y}}$ (b), and $\int -v \frac{\partial}{\partial{y}} (\frac{{p'}}{\rho_0}) dV$ (c) between the original base run (red lines) and the weak-wind run (blue lines). $v$ and $-\frac{1}{\rho_0}\frac{\partial{p'}}{\partial{y}}$ are averaged in the control volume (marked in Fig.~\ref{fig4}d). $\int -v \frac{\partial}{\partial{y}} (\frac{{p'}}{\rho_0}) dV$ is the integration of the horizontal component of $WEF = -v \frac{\partial}{\partial{y}} (\frac{{p'}}{\rho_0}) -w \frac{\partial}{\partial{z}} (\frac{{p'}}{\rho_0})$ in the control volume, and it controls $\int WEF dV$ due to the dominance of horizontal propagation of wave energy. The ellipses highlight the phases where the signal is weakened in the original base run.}
\label{fig_comp_WEF}
\end{figure}

To highlight the nonlinear wave dynamics, the base run from the idealized simulations (discussed in Section 4.a) is rerun with a wind stress that is one order of magnitude weaker ($4\times 10^{-3} N m^{-2}$) and the two solutions are contrasted. Both simulations have inertial oscillations in the across-shore velocity which vary fairly symmetrically over a wave period (Fig.~\ref{fig_comp_WEF}a). In contrast, the across-shore pressure gradient force (PGF) in the original base run exhibits an asymmetric oscillation, as opposed to the regular inertial oscillations in the PGF in the weak-wind run (Fig.~\ref{fig_comp_WEF}b). As a consequence the convergence of wave energy flux does not oscillate at twice the inertial frequency unlike in the weak-wind run  (Fig.~\ref{fig_comp_WEF}c). This suggests that the PGF is asymmetrically modified by the finite amplitude of the waves, and this is what causes the convergence of wave energy flux to pulse inertially rather than semi-inertially. 

\begin{figure}[t]
\centerline{\includegraphics[width=0.85\textwidth]{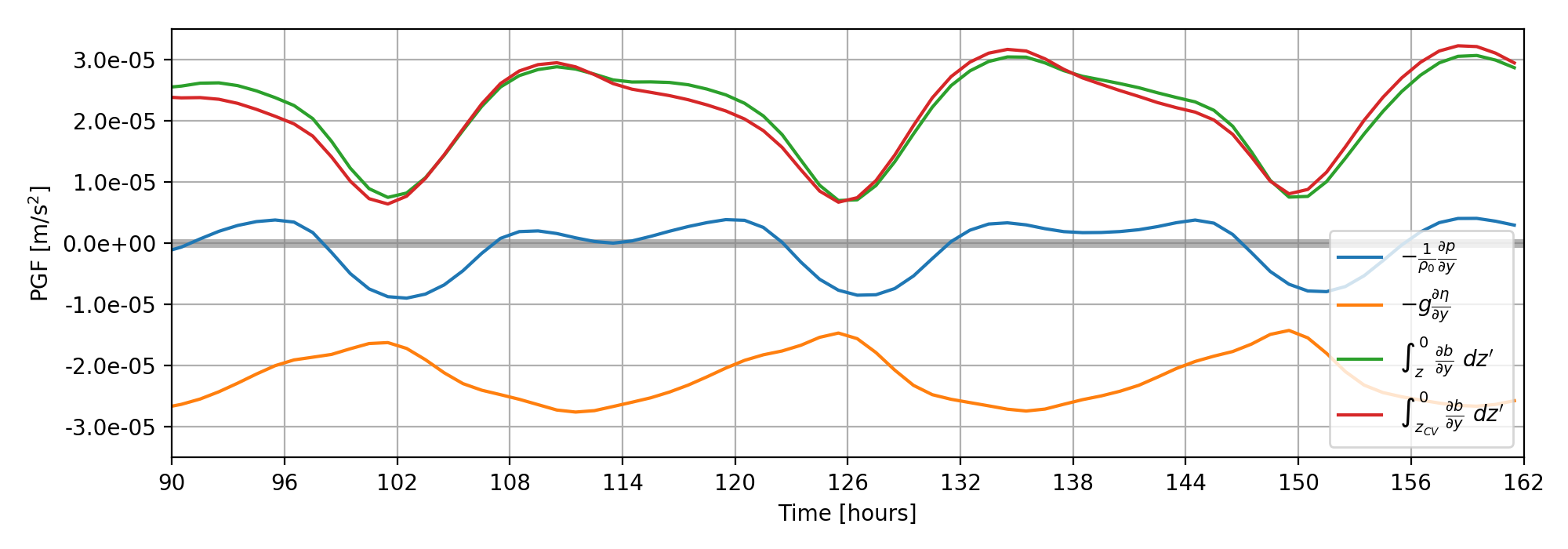}}
\caption{Decomposition of the across-shore PGF $-\frac{1}{\rho_0}\frac{\partial{p}}{\partial{y}}$ (blue line) from the base run. The baroclinic PGF $\int_z^0 \frac{\partial{b}}{\partial{y}} dz'$ is shown in green, and the barotropic PGF $-g\frac{\partial{\eta}}{\partial{y}}$ in red. $\int_{z_{CV}}^0 \frac{\partial{b}}{\partial{y}} dz'$ is the contribution to the baroclinic PGF from above the control volume, where $z_{CV}$ is the vertical location of the top of the control volume. All the PGF terms are averaged in the control volume.
}
\label{fig_decomp_PGF}
\end{figure}

To understand how the waves modify the PGF, the total PGF is decomposed into the barotropic PGF (induced by the sea surface elevation) and the baroclinic PGF (induced by the lateral buoyancy gradient):
\begin{equation}
		-\frac{1}{\rho_0}\frac{\partial{p}}{\partial{y}} = -g\frac{\partial{\eta}}{\partial{y}} + \int_z^0 \frac{\partial{b}}{\partial{y}} dz',
 \label{eq:PGF_decomp}
\end{equation}
where $\eta$ is the sea surface height and $b$ is the buoyancy. This decomposition is conducted in the control volume in the original base run. The baroclinic PGF, $\int_z^0 \frac{\partial{b}}{\partial{y}} dz'$, controls the variability in the total PGF since it sets the phasing of the variations in the total PGF with time (Fig.~\ref{fig_decomp_PGF}). In addition, the variations of the baroclinic PGF (averaged over the control volume) are primarily determined by the lateral buoyancy gradients above the control volume since $\int_{z_{CV}}^0 \frac{\partial{b}}{\partial{y}} dz'$ largely reproduces $\int_z^0 \frac{\partial{b}}{\partial{y}} dz'$ for $z<z_{CV}$ (where $z_{CV}$ is the vertical position of the upper boundary of the control volume). Consequently, the lateral buoyancy gradient above the control volume is the key property of the flow that the waves modify to shape the PGF within the control volume.

\begin{figure}[t]
\centerline{\includegraphics[width=0.8\textwidth]{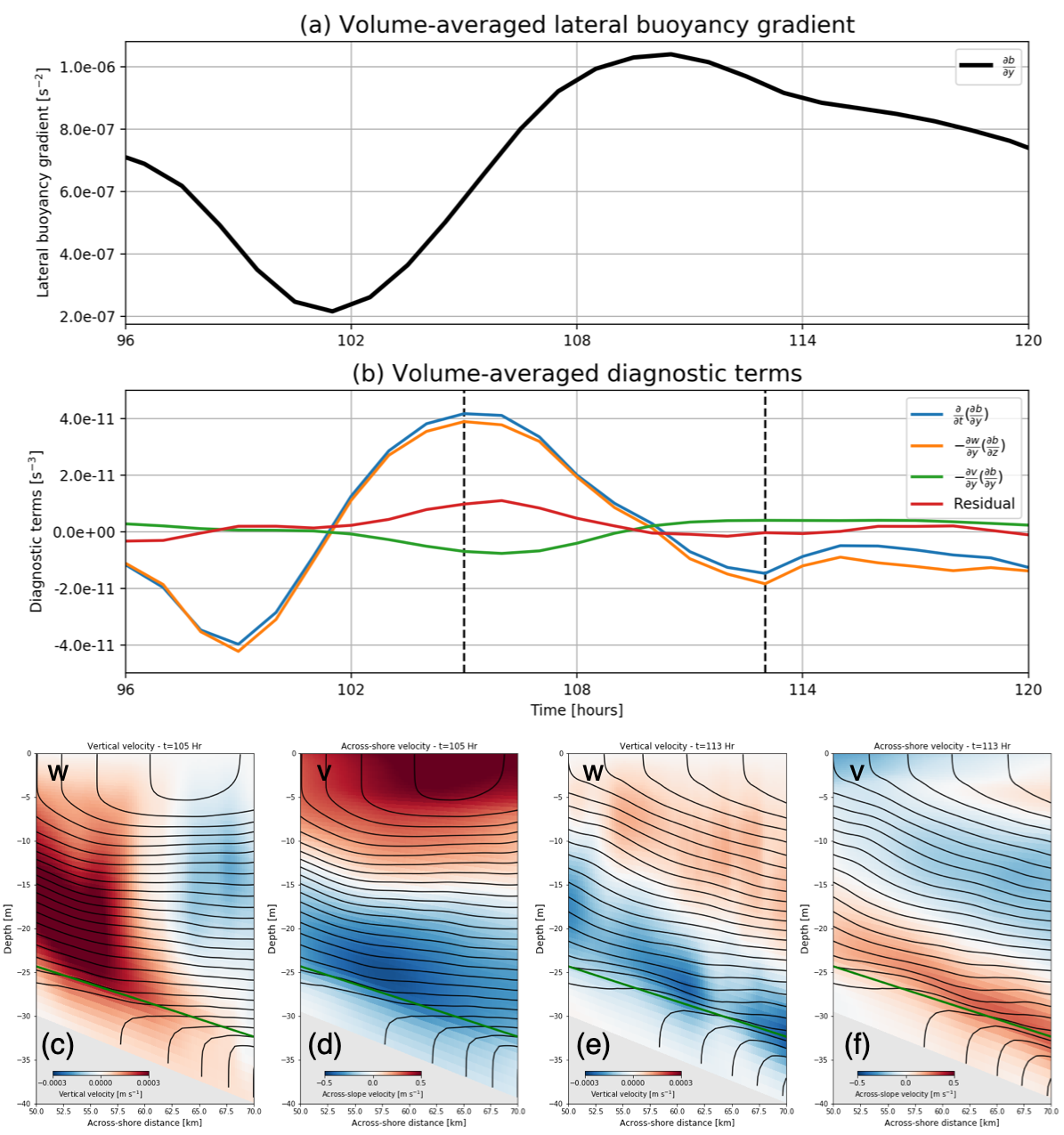}}
\caption{(a) Evolution of the lateral buoyancy gradient over one inertial period from the base run. (b) Time series of the terms governing this evolution, e.g. Eq.~(\ref{eq:M2_diag}). The lateral buoyancy gradient and diagnostic terms are averaged in the region above the control volume (above the green line marked in the lower panels). Differential vertical advection (orange line) primarily controls the rate of change of the lateral buoyancy gradient (blue line). (c and d) Snapshots of w and v at t=105 Hr. The contours are isopycnals plotted every 0.2 $kg\ m^{-3}$. (e and f) Same as c and d but at t=113 Hr.}
\label{fig_M2}
\end{figure}

To further understand how the waves modulate $\frac{\partial{b}}{\partial{y}}$, we diagnose the terms in the equation governing the evolution of this component of the buoyancy gradient:
\begin{equation}
    \frac{D}{Dt}\left( \frac{\partial b}{\partial y} \right) 
        = - \frac{\partial w}{\partial y} \frac{\partial b}{\partial z} - \frac{\partial v}{\partial y} \frac{\partial b}{\partial y} + Residual.
 \label{eq:M2_diag}
\end{equation}
This diagnostic is calculated in the region above the control volume and shows that vertical differential advection, $- \frac{\partial w}{\partial y} \frac{\partial b}{\partial z}$, sets the rate of change of $\frac{\partial{b}}{\partial{y}}$ (Fig.~\ref{fig_M2}b). In addition, $- \frac{\partial w}{\partial y} \frac{\partial b}{\partial z}$ exhibits asymmetric oscillations over an inertial period - strong oscillations in one half of the inertial period and weak oscillations in the other half (Fig.~\ref{fig_M2}b). This is caused by the modification of the background density field by the waves. The finite-amplitude waves periodically modulate the isopycnal slope, creating one phase with relatively flat isopycnals and another with slanted isopycnals. In the phase with relatively flat isopycnals, the waves form a modal structure in the vertical, resembling the vertical structure of waves in the absence of lateral buoyancy gradients. Since isopycnals are nearly flat during the first phase, differential vertical advection is effective and strongly influences the evolution of the lateral buoyancy gradient (Fig.~\ref{fig_M2}b and c). In the other phase, however, isopycnals steepen and run nearly parallel to the bottom, and the wave's velocity runs nearly parallel to isopycnals reducing the efficacy of the waves in advecting buoyancy and changing the lateral buoyancy gradient (Fig.~\ref{fig_M2}b,e,f). In summary, the finite amplitude of the waves allows the background density field to be significantly modified, and drives an asymmetric oscillation in the lateral buoyancy gradient and the PGF. This leads to the inertial pulsing in the convergence of the wave energy flux and hence in the dissipation and mixing.

\section{Conclusions}\label{conclusions_section}
A specific type of NIW critical layer over sloping bathymetry is explored in this study. When isopycnals align with sloping bathymetry in a stratified layer, a critical layer for NIWs with $\omega=f$ forms. Upon entering this critical layer, the waves are trapped and amplified since their group velocity goes to zero, and mixing is enhanced. 

Such slantwise critical layers form in a fully three-dimensional realistic simulation of anticyclonic eddies on the Texas-Louisiana shelf. The realistic simulation exhibits an inertial enhancement of bottom mixing where the energy from surface-generated NIWs is focused in bottom stratified layers on the shelf. Idealized, two-dimensional ROMS simulations reproduce these phenomena, and ray-tracing and analyses of the waves energetics support the idea that the enhanced bottom mixing is caused by the convergence of NIW energy in slantwise critical layers and largely follows this two-dimensional physics. This conclusion is based on results from an ensemble of simulations that cover the relevant parameter space. The ensemble runs show that background flows that more effectively trap wave rays result in stronger wave energy convergence in the critical layer and enhanced mixing. Overall, the link between enhanced mixing and wave trapping is motivated by the realistic simulation, understood using the theoretical analyses, and strengthened by the results from the idealized simulations and ray-tracing solutions.

Although the focus of this study is a particular application on the Texas-Louisiana shelf, the mechanism of NIW amplification in critical layers over sloping bathymetry should be active in other settings. For example, another coastal application could be upwelling systems over continental shelves, where upwelled, dense waters blanket bathymetry. Potential examples include the upwelling systems over the Oregon shelf \citep{federiuk1996model,avicola2007enhanced}, the New Jersey inner shelf \citep{chant2001evolution}, the shelf off of the California coast \citep{nam2013resonant, woodson2007local}, and the Tasmanian shelf, where recent observations suggest evidence of enhanced near-inertial energy and wave trapping in slantwise critical layers \citep{schlosser2019generation}. 
In these upwelling systems, \cite{federiuk1996model} highlight the importance of background flows in modifying the group velocity of NIWs and attribute the observed enhancement of near-inertial energy during periods of upwelling versus downwelling to wave trapping, similar to the mechanism that we have described here. However \cite{federiuk1996model}  did not identified the key criterion for NIW critical layer formation--alignment of isopycnals with bathymetry--that we have determined from our analyses.

Examples of open-ocean flows that can form such critical layers include dense overflows and currents that drive upslope Ekman arrest in bottom boundary layers. One example of the latter is the Kuroshio Current flowing over the continental slope southeast of Kyushu, Japan \citep{nagai2019kuroshio}. Over the shelfbreak, isopycnals tend to align with the bottom suggesting the existence of a slantwise critical layer. In this layer slantwise shear bands suggestive of amplified NIWs are found and coincide with regions where the turbulent dissipation rate is elevated to values of $O(10^{-7}) m^2 s^{-3}$. Another example is the Florida Current on the western side of the Straits of Florida. On this side of the Straits, the alignment between the isopycnals and the continental slope indicates a slantwise critical layer, where in fact observations show that turbulence can be enhanced in stratified layers off the bottom \citep{winkel02}. We plan to study the dynamics of these open-ocean NIW critical layers in future work.

Diapycnal transport within these critical layers can also be enhanced due to turbulence driven by wave trapping.  Such diapycnal transport can influence the distribution of biogeochemical tracers such as iron and oxygen and thus potentially influence coastal ecosystems. In the open ocean, NIW trapping in critical layers could affect abyssal diapycnal transport near the bottom, which could modify water mass distributions and influence the meridional overturning circulation.

\clearpage
\acknowledgments
This work was funded by the SUNRISE project, NSF grant numbers OCE-1851450 (L.Q. and L.N.T) and OCE-1851470 (R.H.D). We thank Olivier Asselin, Bertrand Delorme, Jinliang Liu, Jen MacKinnon, Jonathan Nash, Guillaume Roullet, Kipp Shearman, and John Taylor for very helpful suggestions when preparing this manuscript.

\clearpage
\appendix[A]
\appendixtitle{Rotated and non-rotated coordinates}
The rotated coordinates are rotated with the angle of $\theta$ (considered as positive) to align with the sloping topography. The relation between the non-rotated and rotated coordinates is
\begin{equation}
 \begin{aligned}
    \Tilde{y} &= \ cos \theta y - sin \theta z, \\
    \Tilde{z} &= sin \theta y + cos \theta z, \\
 \end{aligned}
 \label{Deq:relation1}
\end{equation}
where $\Tilde{y}$ denotes the rotated, across-slope direction and $\Tilde{z}$ denotes the rotated, slope-normal direction. Assuming that the isopycnals are parallel with the sloping bathymetry, the buoyancy field has the slope-normal gradient as a function only depending on $\Tilde{z}$ (i.e., $\Tilde{N}^2 (\Tilde{z}) \equiv \frac{\partial  \overline{b}}{\partial \Tilde{z}} (\Tilde{z})$) and has no across-slope gradient (i.e., $\frac{\partial  \overline{b}}{\partial \Tilde{y}}=0$). In the non-rotated coordinates, the horizontal and vertical buoyancy gradients can be linked with $\Tilde{N}^2$ by
\begin{equation}
 \begin{aligned}
    M^2 & \equiv \frac{\partial  \overline{b}}{\partial y} = \frac{\partial  \overline{b}}{\partial \Tilde{y}} \frac{\partial \Tilde{y}}{\partial y} + \frac{\partial  \overline{b}}{\partial \Tilde{z}} \frac{\partial \Tilde{z}}{\partial y} = \Tilde{N}^2 sin \theta \ ,\\
    N^2 & \equiv \frac{\partial  \overline{b}}{\partial z} = \frac{\partial  \overline{b}}{\partial \Tilde{y}} \frac{\partial \Tilde{y}}{\partial z} + \frac{\partial  \overline{b}}{\partial \Tilde{z}} \frac{\partial \Tilde{z}}{\partial z} = \Tilde{N}^2 cos \theta \ .\\
 \end{aligned}
 \label{Deq:Bgrad}
\end{equation}
Similarly, the horizontal and vertical gradients of the background along-slope velocity $ \overline{u}$ are
\begin{equation}
 \begin{aligned}
    \frac{\partial  \overline{u}}{\partial y} & =\frac{\partial  \overline{u}}{\partial \Tilde{z}} sin \theta \ ,\\
    \frac{\partial  \overline{u}}{\partial z} & =\frac{\partial  \overline{u}}{\partial \Tilde{z}} cos \theta \ .\\
 \end{aligned}
 \label{Deq:ugrad}
\end{equation}
Noting that $ \overline{u}$ is in the thermal wind balance with the background buoyancy, the vertical shear of $ \overline{u}$ can be obtained, by using Eq.(\ref{Deq:Bgrad}), as
\begin{equation}
 \begin{aligned}
    \frac{\partial  \overline{u}}{\partial z} = - \frac{1}{f} \frac{\partial  \overline{b}}{\partial y} =- \frac{\Tilde{N}^2}{f} sin \theta \ .\\
 \end{aligned}
 \label{Deq:TW}
\end{equation}
Given Eq.(\ref{Deq:ugrad}), the horizontal gradient of $ \overline{u}$ can be then written as
\begin{equation}
 \begin{aligned}
    \frac{\partial  \overline{u}}{\partial y} = \frac{\partial  \overline{u}}{\partial z} tan \theta =- \frac{\Tilde{N}^2}{f} sin \theta\ tan \theta \ .\\
 \end{aligned}
 \label{Deq:uhshear}
\end{equation}
Consequently, the vorticity Rossby number $Ro_g$ and Richardson number $Ri_g$ can be expressed as
\begin{equation}
 \begin{aligned}
    Ro_g &= -\frac{\partial  \overline{u}}{\partial y} /f = \frac{\Tilde{N}^2}{f^2} sin \theta\ tan \theta  \ ,\\
    Ri_g &= \frac{\partial  \overline{b}}{\partial z} / (\frac{\partial  \overline{u}}{\partial z})^2 = \frac{f^2}{\Tilde{N}^2 sin \theta\ tan \theta}   \ .\\
 \end{aligned}
 \label{Deq:RiRo}
\end{equation}

\clearpage
\appendix[B]
\appendixtitle{Ray tracing}
Rays are calculated by integrating the following equation
\begin{equation}
    \frac{dz^r}{dy^r}=s_{ray}=s_{\rho}\pm \sqrt{\frac{\omega^2-\omega^2_{min}}{N^2}} 
 \label{eq:slope}
\end{equation}
where $(y^r,z^r)$ is the position of the ray in the $y-z$ plane.  At a certain discrete location of the path $(y^r_{n},z^r_{n})$, it is possible to calculate $s_{\rho}$, $\omega_{min}$, and $N^2$, thus, the slope of the path $s_{ray}$ can be obtained for a wave of frequency $\omega$. With a small change in $y$, $\delta y^r=y^r_{n+1}-y^r_{n}$, the next vertical location of the ray $z^r_{n+1}$ can be estimated as 
\begin{equation}
\begin{aligned}
    z^r_{n+1} &= \int_{y^r_n}^{y^r_{n+1}} s_{ray}\ dy^r + z^r_n \approx  s_{ray} |_{(y^r_n,z^r_n)} \delta y^r + z^r_n.
\end{aligned}
 \label{eq:zn}
\end{equation}
Starting at an initial point, recursively calculating Eq.~\ref{eq:slope} and \ref{eq:zn} will trace out the path of a wave packet.

\clearpage
\bibliographystyle{ametsoc2014}
\bibliography{references}


\begin{table}
\caption{Parameters used in the base run (first row) and ensemble runs (second row). $\alpha$ is bottom slope. $f$ is Coriolis parameter. $|\vec{\tau}|$ is the amplitude of the oscillatory, across-slope wind stress. $L_C$, $L_T$, and $L$ are the length scales of the critical layer, transition zone, and offshore front, respectively. $\zeta_0$ is the surface relative vorticity of the offshore front. $N^2$ is the stratification in the non-rotated coordinates. Only $L$, $\zeta_0$, and $N^2$ vary in the ensemble simulations, and there are a total of 18 ensemble runs.}

\begin{tabular*}{\hsize}{@{\extracolsep\fill}lcccccccc@{}}
\topline
$f$ (s$^{-1}$) & $\alpha$ &  $|\vec{\tau}|$ (N m$^{-2}$) & $L_C$ (km) & $L_T$ (km) & $L$ (km) & $\zeta_0$  & $N^2$ (s$^{-2}$)\\
\midline
7.27e-05 & 5.00e-04 &  4.00e-02 &  50.0 & 20.0 &  40.0 & -0.3f & 3.00e-03\\
-        & -        &   -    &  -    &  -  &  (30.0, 40.0, 50.0) & (-0.3f, -0.5f, -0.7f) & (3.00e-03, 5.00e-03)\\

\botline
\label{tab1}
\end{tabular*}
\end{table}


\end{document}